\documentclass{ptephy}

\usepackage{amsmath} 
\usepackage{amsthm} 
\usepackage{amsfonts}
\usepackage{latexsym}
\usepackage{amssymb} 
\usepackage{verbatim}

\newcommand{\RF}{{{\mathbb R}}}

\newtheorem{conjecture}{Conjecture}

\begin{document}

\title{
  Construction of gauge-invariant variables of linear metric
  perturbations on an arbitrary background spacetime 
}

\author{
  \name{Kouji Nakamura}{1}
}

\address{
  \affil{1}{
    TAMA Project, Optical and Infrared Astronomy Division,\\
    National Astronomical Observatory of Japan,\\
    2-21-1, Osawa, Mitaka, Tokyo 181-8588, Japan
  }
  \email{kouji.nakamura@nao.ac.jp}
}
\begin{abstract}%
  An outline of a proof of the decomposition of linear metric
  perturbations into gauge-invariant and gauge-variant parts on
  the an arbitrary background spacetime which admits ADM
  decomposition is discussed.
  We explicitly construct the gauge-invariant and gauge-variant
  parts of the linear metric perturbations through the
  assumption of the existence of some Green functions.
  We also confirm the result through another approach.
  This implies that we can develop the higher-order
  gauge-invariant perturbation theory on an arbitrary background
  spacetime.
  Remaining issues to complete the general-framework of the
  general-relativistic higher-order gauge-invariant perturbation
  theories are also discussed.
\end{abstract}
\subjectindex{E01}
\maketitle

\section{Introduction}
\label{sec:intro}


Perturbation theories are powerful techniques in many area of
physics and the developments of perturbation theories lead
physically fruitful results and interpretations of natural
phenomena.


In physics, researchers want to describe realistic situations in
a compact manner. 
Exact solutions in a theory for natural phenomena are candidates
which can describe realistic situations.  
However, in many cases, realistic situations are too complicated
and often difficult to describe by an exact solution of a
theory. 
This difficulty may be due to the fact that exact solutions only
describe special cases even if the theory is appropriate to
describe the natural phenomena, or may be due to the lack of the
applicability of the theory itself.
Even in the case where an exact solution of a theory well
describes a physical situation, the properties of the natural 
system will not be completely described only through the exact
solution. 
In natural phenomena, there always exist ``fluctuations''.
In this case, perturbative treatments of the theory is a
powerful tool and researchers investigate perturbative approach 
within a theory to clarify the properties of fluctuations.


General relativity is one of theories in which the construction
of exact solutions is not so easy.
Although there are many exact solutions to the Einstein
equation~\cite{H.Stephani-D.Kramer-M.A.N.MacCallum-C.Hoenselaers-E.Herlt-2003}, 
these are often too idealized.
Of course, there are some exact solutions to the Einstein
equation which well-describe our universe, or gravitational
field of stars and black holes.
These exact solutions by itself do not describe fluctuations
around these exact solutions.
To describe them, we have to consider the perturbations.
Therefore, general relativistic {\it linear} perturbation theory
is a useful technique to investigate the properties of
fluctuations around exact
solutions~\cite{Bardeen-1980,Gerlach_Sengupta-1979,Kodama-Ishibashi-2004}. 
Through these linear perturbation theories, we can describe
fluctuations such as the density or the temperature fluctuations
of our universe, gravitational waves from self-gravitating
objects.


On the other hand, {\it higher-order} general-relativistic
perturbations also have very wide applications.
Among these applications, second-order cosmological
perturbations are topical
subject~\cite{Tomita-1967-Non-Gaussianity,M.Bruni-S.Soonego-CQG1997,S.Sonego-M.Bruni-CMP1998,Nakamura:2010yg,kouchan-cosmo-second}
due to the precise measurements in recent
cosmology~\cite{Non-Gaussianity-observation}. 
Higher-order black hole perturbations are also discussed in some
literature~\cite{Gleiser-Nicasio-2000}.
Moreover, as a special example of higher-order perturbation
theory, there are researches on perturbations of a spherical
star~\cite{Kojima-1997}, which are motivated by the
investigation of the oscillatory behaviors of a rotating neutron
star.
Thus, there are many physical situations to which general
relativistic higher-order perturbation theory should be
applied.


As well-known, general relativity is based on the concept of
general covariance.
Intuitively speaking, the principle of general covariance states
that there is no preferred coordinate system in nature, and the
notion of general covariance is mathematically included in the
definition of a spacetime manifold in a trivial way.
This is based on the philosophy that coordinate systems are
originally chosen by us, and that natural phenomena have nothing
to do with our coordinate system.
Due to this general covariance, the ``gauge degree of freedom'', 
which is unphysical degree of freedom of perturbations, arises
in general-relativistic perturbations.
To obtain physical results, we have to fix this gauge degrees
of freedom or to extract some invariant quantities of
perturbations.
This situation becomes more complicated in higher-order
perturbation theory.
In some linear perturbation theories on some background
spacetimes, there are so-called {\it gauge-invariant}
perturbation theories.
In these theories, one may treat only variables which are
independent of gauge degree of freedom without any gauge
fixing.
Therefore, it is worthwhile to investigate higher-order
gauge-invariant perturbation theory from a general point of
view.


According to these motivations, the general framework of
higher-order general-relativistic gauge-invariant perturbation
theory has been discussed in some
papers~\cite{kouchan-gauge-inv,kouchan-second} by the present
author.
Although these development of higher-order perturbation theory
was originally motivated by researches on the oscillatory
behaviors of a self-gravitating Nambu-Goto
membrane~\cite{kouchan-papers}, these works are applicable to
cosmological perturbations and we clarified the gauge-invariance
of the second-order perturbations of the Einstein
equations~\cite{kouchan-cosmo-second,kouchan-second-cosmo-matter,kouchan-second-cosmo-consistency}.


In Ref.~\cite{kouchan-gauge-inv}, we proposed a procedure to
find gauge-invariant variables for higher-order perturbations on 
an arbitrary background spacetime.
This proposal is based on the single assumption that {\it we
  already know the procedure to find gauge-invariant variables
  for linear-order metric perturbations}
(Conjecture \ref{conjecture:decomposition-conjecture} in
Sec.~\ref{sec:General-framework-of-the-gauge-invariant-perturbation-theory}
in this paper).
Under the same assumption, we summarize some formulae for the
second-order perturbations of the curvatures and energy-momentum
tensor for the matter fields in
Refs.~\cite{kouchan-second,kouchan-second-cosmo-matter}.
Confirming that the above assumption in the case of cosmological
perturbations is correct, in Refs.~\cite{kouchan-cosmo-second},
we develop the second-order gauge-invariant cosmological
perturbation theory.
Through these works, we find that our general framework of
higher-order gauge-invariant perturbation theory is well-defined
except for the above assumption for linear-order metric
perturbations.
Therefore, we proposed the above assumption as a conjecture in
Ref.~\cite{kouchan-second-cosmo-matter}.
If this conjecture is true, the higher-order
general-relativistic gauge-invariant perturbation theory is
completely formulated on an arbitrary background spacetime and
has very wide applications.


The main purpose of this paper is to give a scenario of a proof
of this conjecture based on the premise that the background
spacetime admits ADM decomposition.
We explicitly construct the gauge-invariant and gauge-variant
parts of the linear metric perturbation.
Although some special modes are excluded in the proof in this
paper, we may say that the above conjecture is almost correct
for linear-order perturbations on an arbitrary background
spacetime.
This paper is the full paper version of our previous short
letter~\cite{kouchan-decomp-letter-version}.


The organization of this paper is as follows.
In
Sec.~\ref{sec:General-framework-of-the-gauge-invariant-perturbation-theory}, 
we briefly review the general framework of the second-order
gauge-invariant perturbation theory developed in
Refs.~\cite{kouchan-gauge-inv,kouchan-second} and the above
conjecture is also declared as Conjecture
\ref{conjecture:decomposition-conjecture} in this review. 
In
Sec.~\ref{sec:K.Nakamura-2010-3}, we give a scenario of a proof
of Conjecture \ref{conjecture:decomposition-conjecture}.
We note that we assume that the existence of Green functions 
for two elliptic differential operators in our outline of a
proof.
Therefore, the modes which belong to the kernel of these two
elliptic differential operators are excluded in this paper.
Since we use tricky logic in our outline in
Sec.~\ref{sec:K.Nakamura-2010-3}, we reconsider the derivation 
of gauge-transformation rules through an alternative approaches
in Sec.~\ref{sec:K.Nakamura-2010-4} to check the consistency of
our result in Sec.~\ref{sec:K.Nakamura-2010-3}.
The final section (Sec.~\ref{sec:summary}) is devoted to summary
and discussions.
Throughout this paper, we use the covariant decomposition of
symmetric tensors on Riemannian manifold which was developed by
York~\cite{J.W.York-1973-1974}. 
We review York's discussions on this covariant decomposition 
in
Appendix~\ref{sec:S.Deser-1967-J.W.York.Jr-1973-J.W.York.Jr-1974}.
Although his discussions are for tensors on a closed manifold,
we use his decomposition on a finite region in a Riemannian
manifold with boundaries assuming the existence of Green
functions for two elliptic differential operators.


We employ the notation in
Refs.~\cite{kouchan-gauge-inv,kouchan-second} and use abstract
index notation~\cite{Wald-book}.
We also employ natural units in which the velocity of light
satisfies $c=1$.


\section{General framework of the higher-order gauge-invariant perturbation theory}
\label{sec:General-framework-of-the-gauge-invariant-perturbation-theory}


In this section, we briefly review the general framework of the 
gauge-invariant perturbation theory developed in
Ref.~\cite{kouchan-gauge-inv}.
The aim of this section is to emphasize that Conjecture 
\ref{conjecture:decomposition-conjecture} is an important
premise of our general framework.
In Sec.~\ref{sec:Notion_of_gauge_in_general_relativity}, we review
the notion of the {\it gauge} in general relativity and
{\it gauge degree of freedom} in general-relativistic
perturbations.
In Sec.~\ref{sec:Perturbations_in_general_relativity}, the
definition of perturbations in general relativity and its 
gauge-transformation rules are reviewed.
When we consider perturbations in any theory with general
covariance, we have to exclude these gauge degrees of freedom in
the perturbations.
To accomplish this, {\it gauge-invariant variables} of
perturbations are useful, and these are regarded as physically
meaningful quantities.
In Sec.~\ref{sec:gauge-invariant-variables}, a procedure to find
gauge-invariant variables of perturbations is explained, which
was developed in Ref.~\cite{kouchan-gauge-inv}.
We emphasize that the ingredients of this section do not
depend on the details of the background spacetime, except for
Conjecture \ref{conjecture:decomposition-conjecture}.


\subsection{Notion of ``gauge'' in general relativity}
\label{sec:Notion_of_gauge_in_general_relativity}


General relativity is a theory based on general covariance,
which intuitively states that there is no preferred coordinate
system in nature.
Due to this general covariance, the notion of ``gauge'' is
introduced in the theory.
Sachs~\cite{R.K.Sachs-1964} is the first person who pointed out
that there are two kinds of ``gauges'' in general relativity.
He called these two ``gauges'' as the first- and the second-kind
of gauges, respectively.
The distinction of these two different notion of ``gauges'' is
an important premise in the arguments in
Sec.~\ref{sec:K.Nakamura-2010-3}.
Therefore, first of all, we remind the difference of these two
``gauges''.


{\it The first kind gauge} is a coordinate system on a single
manifold ${\mathcal M}$.
On a manifold, we can always introduce a coordinate system as a
diffeomorphism $\psi_{\alpha}$ which maps from an open set
$O_{\alpha}\subset{\mathcal M}$ to
$\psi_{\alpha}(O_{\alpha})\subset\RF^{n+1}$ (where
$n+1=\dim{\mathcal M}$).
This coordinate system $\psi_{\alpha}$ is called
{\it gauge choice of the first kind}.
If we consider another open set $O_{\beta}\subset{\mathcal M}$, we
have another gauge choice $\psi_{\beta}$ : 
$O_{\beta}\mapsto\psi_{\beta}(O_{\beta})\subset\RF^{n+1}$.
If $O_{\alpha}\cap O_{\beta}\neq\emptyset$, we can consider the
diffeomorphism $\psi_{\beta}\circ\psi_{\alpha}^{-1}$, which is a
coordinate transformation : 
$\psi_{\alpha}(O_{\alpha}\cap O_{\beta})\subset\RF^{n+1}$
$\mapsto$ $\psi_{\beta}(O_{\alpha}\cap O_{\beta})\subset\RF^{n+1}$.
This coordinate transformation
$\psi_{\beta}\circ\psi_{\alpha}^{-1}$ is also called
{\it gauge transformation of the first kind} in general
relativity.


On the other hand, {\it the second kind gauge} appears in
perturbation theories in a theory with general covariance.
This is the main issue of this paper.
In perturbation theories, we always treat two spacetime
manifolds.
One is the physical spacetime ${\mathcal M}$ which is our nature
itself and we want to clarify the properties of ${\mathcal M}$
through perturbations.
Another is the background spacetime ${\mathcal M}_{0}$ which has
nothing to do with our nature but is prepared by hand for
perturbative analyses.
Let us denote the physical spacetime by $({\mathcal M},\bar{Q})$
and the background spacetime by $({\mathcal M}_{0},Q_{0})$, where 
$\bar{Q}$ is the collection of tensor fields on ${\mathcal M}$, and
$Q_{0}$ is the collection of the background values on 
${\mathcal M}_{0}$ for the collection $\bar{Q}$ on ${\mathcal M}$.
{\it The gauge choice of the second kind} is the point
identification map ${\mathcal X}$ $:$ 
${\mathcal M}_{0}\mapsto{\mathcal M}$~\cite{R.K.Sachs-1964,J.M.Stewart-M.Walker11974}. 
We have to note that the correspondence ${\mathcal X}$ between
points on ${\mathcal M}_{0}$ and ${\mathcal M}$ is not unique to the
perturbation theory with general covariance.
General covariance intuitively means that there is no
preferred coordinate system in the theory.
Due to general covariance, we have no guiding principle to
choose the identification map ${\mathcal X}$.
Actually, as a gauge choice of the second kind, we may choose a
different point identification map ${\mathcal Y}$ from ${\mathcal X}$.
This implies that there is degree of freedom in the gauge choice
of the second kind.
This is {\it the gauge degree of freedom of the second kind} in
perturbation theory.
In this understanding, 
{\it the gauge transformation of the second kind} is a change
${\mathcal X}\rightarrow{\mathcal Y}$ of the identification map 
${\mathcal M}_{0}\mapsto{\mathcal M}$.


\subsection{Perturbations in general relativity}
\label{sec:Perturbations_in_general_relativity}


To formulate the above second-kind gauge in more detail, we
introduce an infinitesimal parameter $\lambda$ for
perturbations.
Further, we consider the $(n+1)+1$-dimensional manifold 
${\mathcal N}={\mathcal M}\times\RF$, where $n+1=\dim{\mathcal M}$ and
$\lambda\in\RF$.
The background spacetime 
${\mathcal M}_{0}=\left.{\mathcal N}\right|_{\lambda=0}$ and the
physical spacetime
${\mathcal M}={\mathcal M}_{\lambda}=\left.{\mathcal N}\right|_{\RF=\lambda}$ are
also submanifolds embedded in the extended manifold ${\mathcal N}$.
Each point on ${\mathcal N}$ is identified by a pair, $(p,\lambda)$,
where $p\in{\mathcal M}_{\lambda}$, and each point in the background
spacetime ${\mathcal M}_{0}$ in ${\mathcal N}$ is identified by
$\lambda=0$.
Through this construction, the manifold ${\mathcal N}$ is foliated by
$(n+1)$-dimensional submanifolds ${\mathcal M}_{\lambda}$ of each
$\lambda$, and these are diffeomorphic to the physical
spacetime ${\mathcal M}$ and the background spacetime
${\mathcal M}_{0}$.
The manifold ${\mathcal N}$ has a natural differentiable structure
consisting of the direct product of ${\mathcal M}$ and $\RF$.
Further, the perturbed spacetimes ${\mathcal M}_{\lambda}$ for each
$\lambda$ must have the same differential structure by this
construction.


If a tensor field $Q_{\lambda}$ is given on each
${\mathcal M}_{\lambda}$, $Q_{\lambda}$ is automatically extended to a
tensor field on ${\mathcal N}$ by $Q(p,\lambda):=Q_{\lambda}(p)$,
where $p\in{\mathcal M}_{\lambda}$.
Tensor fields on ${\mathcal N}$ obtained through this construction
are necessarily ``tangent'' to each ${\mathcal M}_{\lambda}$, i.e.,
their normal component to each ${\mathcal M}_{\lambda}$ in
${\mathcal N}$ identically vanishes.
To consider the basis of the tangent space of ${\mathcal N}$, we
introduce the normal form $(d\lambda)_{a}$ and its dual
$(\partial/\partial\lambda)^{a}$, which are normal to each
${\mathcal M}_{\lambda}$ in ${\mathcal N}$.
These satisfy $(d\lambda)_{a}
\left(\partial/\partial\lambda\right)^{a} = 1$.
$(d\lambda)_{a}$ and $(\partial/\partial\lambda)^{a}$ are normal
to any tensor field extended from the tangent space on each
${\mathcal M}_{\lambda}$ through the above construction. 
The set consisting of $(d\lambda)_{a}$,
$(\partial/\partial\lambda)^{a}$, and the basis of the tangent 
space on each ${\mathcal M}_{\lambda}$ is regarded as the basis of
the tangent space of ${\mathcal N}$.


To define perturbations of an arbitrary tensor field $\bar{Q}$,
we have to compare $\bar{Q}$ on the physical spacetime
${\mathcal M}_{\lambda}$ with $Q_{0}$ on the background spacetime
${\mathcal M}_{0}$ through the introduction of a gauge choice of the
second kind.
The gauge choice of the second kind is made by assigning a
diffeomorphism ${\mathcal X}_{\lambda}$ $:$ ${\mathcal N}$ $\rightarrow$
${\mathcal N}$ such that ${\mathcal X}_{\lambda}$ $:$ ${\mathcal M}_{0}$
$\rightarrow$ ${\mathcal M}_{\lambda}$.
The pull-back ${\mathcal X}_{\lambda}^{*}$, which is induced by the
map ${\mathcal X}_{\lambda}$, maps a tensor field $\bar{Q}$ on 
${\mathcal M}_{\lambda}$ to a tensor field 
${\mathcal X}_{\lambda}^{*}\bar{Q}$ on ${\mathcal M}_{0}$. 
Once the definition of the pull-back of the gauge choice 
${\mathcal X}_{\lambda}$ is given, the perturbations of a tensor
field $\bar{Q}$ under the gauge choice ${\mathcal X}_{\lambda}$ are
simply defined by the evaluation of the Taylor expansion at 
${\mathcal M}_{0}$ in ${\mathcal N}$:
\begin{equation}
  \label{eq:Bruni-35}
  {}^{\mathcal X}\!Q
  :=
  \left.{\mathcal X}^{*}_{\lambda}\bar{Q}_{\lambda}\right|_{{\mathcal M}_{0}} 
  =
  Q_{0}
  + \lambda {}^{(1)}_{\;\mathcal X}\!Q
  + \frac{1}{2} \lambda^{2} {}^{(2)}_{\;\mathcal X}\!Q
  + O(\lambda^{3}),
\end{equation}
where ${}^{(1)}_{\;\mathcal X}\!Q$ and ${}^{(2)}_{\;\mathcal X}\!Q$ are
the first- and the second-order perturbations of $\bar{Q}$,
respectively.


We also note that these perturbations completely depend on the
gauge choice ${\mathcal X}_{\lambda}$.
When we have two different gauge choices ${\mathcal X}_{\lambda}$
and ${\mathcal Y}_{\lambda}$, we have two different representations
of the perturbative expansion of the pulled-backed variables 
$\left.{\mathcal X}^{*}_{\lambda}\bar{Q}_{\lambda}\right|_{{\mathcal M}_{0}}$
in Eq.~(\ref{eq:Bruni-35}) and 
$\left.{\mathcal Y}^{*}_{\lambda}\bar{Q}_{\lambda}\right|_{{\mathcal M}_{0}}$:
\begin{eqnarray}
  \label{eq:Ygaueg-perturbation}
  {}^{\mathcal Y}\!Q &:=&
  \left.{\mathcal Y}^{*}_{\lambda}Q_{\lambda}\right|_{{\mathcal M}_{0}}
  =
  Q_{0}
  + \lambda {}^{(1)}_{\;\mathcal Y}\!Q
  + \frac{1}{2} \lambda^{2} {}^{(2)}_{\;\mathcal Y}\!Q
  + O(\lambda^{3}),
\end{eqnarray}
Although these two representations of the perturbations are
different from each other, these should be equivalent because of
general covariance.
This equivalence is guaranteed by the 
{\it gauge-transformation rules} between two different gauge
choices.
The change of the gauge choice from ${\mathcal X}_{\lambda}$ to
${\mathcal Y}_{\lambda}$ is represented by the diffeomorphism 
$\Phi_{\lambda}:=({\mathcal X}_{\lambda})^{-1}\circ{\mathcal Y}_{\lambda}$.
This diffeomorphism $\Phi_{\lambda}$ is the map $\Phi_{\lambda}$
$:$ ${\mathcal M}_{0}$ $\rightarrow$ ${\mathcal M}_{0}$ for each value
of $\lambda\in\RF$ and does change the point identification.
Therefore, the diffeomorphism $\Phi_{\lambda}$ is regarded as
the gauge transformation $\Phi_{\lambda}$ $:$
${\mathcal X}_{\lambda}$ $\rightarrow$ ${\mathcal Y}_{\lambda}$.
The gauge transformation $\Phi_{\lambda}$ induces a pull-back
from the representation ${}^{\mathcal X}Q_{\lambda}$ of the
perturbed tensor field $Q$ in the gauge choice 
${\mathcal X}_{\lambda}$ to the representation 
${}^{\mathcal Y}Q_{\lambda}$ in the gauge choice 
${\mathcal Y}_{\lambda}$.
Actually, the tensor fields ${}^{\mathcal X}Q_{\lambda}$ and
${}^{\mathcal Y}Q_{\lambda}$, which are defined on ${\mathcal M}_{0}$,
are connected by the linear map $\Phi^{*}_{\lambda}$ as
${}^{\mathcal Y}Q_{\lambda}=\Phi^{*}_{\lambda} {}^{\mathcal X}Q_{\lambda}$.
According to generic arguments concerning the Taylor expansion
of the pull-back of tensor fields on the same
manifold~\cite{S.Sonego-M.Bruni-CMP1998,Nakamura:2010yg}, it
should be expressed the gauge transformation 
$\Phi^{*}_{\lambda} {}^{\mathcal X}Q_{\lambda}$ in the form
\begin{eqnarray}
  {}^{\mathcal Y}\!Q =
  \Phi^{*}_{\lambda}{}^{\mathcal X}\!Q = {}^{\mathcal X}\!Q
  + \lambda {\pounds}_{\xi_{(1)}} {}^{\mathcal X}\!Q
  + \frac{\lambda^{2}}{2} \left\{
    {\pounds}_{\xi_{(2)}} + {\pounds}_{\xi_{(1)}}^{2}
  \right\} {}^{\mathcal X}\!Q
  + O(\lambda^{3}),
  \label{eq:Bruni-46-one} 
\end{eqnarray}
where the vector fields $\xi_{(1)}^{a}$ and $\xi_{(2)}^{a}$ are
the generators of the gauge transformation $\Phi_{\lambda}$.
Substituting Eqs.~(\ref{eq:Bruni-35}) and
(\ref{eq:Ygaueg-perturbation}) into Eq.~(\ref{eq:Bruni-46-one}),
we obtain the gauge-transformation rules for perturbations
${}^{(1)}_{\;{\mathcal X}}\!Q$ and ${}^{(2)}_{\;\mathcal X}\!Q$ as follows:
\begin{eqnarray}
  \label{eq:Bruni-47-one}
  {}^{(1)}_{\;{\mathcal Y}}\!Q - {}^{(1)}_{\;{\mathcal X}}\!Q &=& 
  {\pounds}_{\xi_{(1)}}Q_{0}, \\
  \label{eq:Bruni-49-one}
  {}^{(2)}_{\;\mathcal Y}\!Q - {}^{(2)}_{\;\mathcal X}\!Q &=& 
  2 {\pounds}_{\xi_{(1)}} {}^{(1)}_{\;\mathcal X}\!Q 
  +\left\{{\pounds}_{\xi_{(2)}}+{\pounds}_{\xi_{(1)}}^{2}\right\} Q_{0}.
\end{eqnarray}


The notion of gauge invariance considered in this paper is the  
{\it order by order gauge invariance} proposed in
Ref.~\cite{kouchan-second-cosmo-matter}. 
We call the $k$th-order perturbation ${}^{(k)}_{{\mathcal X}}\!Q$ is
gauge invariant iff 
\begin{equation}
  {}^{(k)}_{\;\mathcal X}\!Q = {}^{(k)}_{\;\mathcal Y}\!Q
\end{equation}
for any gauge choice ${\mathcal X}_{\lambda}$ and
${\mathcal Y}_{\lambda}$. 
Through this concept of the order by order gauge invariance, we
can decompose any perturbation of $Q$ into the gauge-invariant
and gauge-variant parts, as shown in Ref.~\cite{kouchan-gauge-inv}. 
In terms of these gauge-invariant variables, we can develop the
gauge-invariant perturbation theory.
However, this development is based on a non-trivial conjecture,
i.e., Conjecture \ref{conjecture:decomposition-conjecture} for
linear order metric perturbations as explained below.


\subsection{Gauge-invariant variables}
\label{sec:gauge-invariant-variables}


Inspecting the gauge-transformation rules
(\ref{eq:Bruni-47-one}) and (\ref{eq:Bruni-49-one}), we define
gauge-invariant variables for metric perturbations and for
perturbations of an arbitrary matter field.
First, we consider the metric perturbation.
The metric $\bar{g}_{ab}$ on ${\mathcal M}$, which is pulled back to 
${\mathcal M}_{0}$ using a gauge choice ${\mathcal X}_{\lambda}$, is
expanded in the form of Eq.~(\ref{eq:Bruni-35}):
\begin{eqnarray}
  {\mathcal X}^{*}_{\lambda}\bar{g}_{ab}
  &=&
  g_{ab} + \lambda {}_{{\mathcal X}}\!h_{ab} 
  + \frac{\lambda^{2}}{2} {}_{{\mathcal X}}\!l_{ab}
  + O^{3}(\lambda),
  \label{eq:metric-expansion}
\end{eqnarray}
where $g_{ab}$ is the metric on ${\mathcal M}_{0}$.
Of course, the expansion (\ref{eq:metric-expansion}) of the
metric depends entirely on the gauge choice 
${\mathcal X}_{\lambda}$.
Nevertheless, henceforth, we do not explicitly express the index
of the gauge choice ${\mathcal X}_{\lambda}$ if there is no
possibility of confusion.


Based on these setups, in Ref.~\cite{kouchan-gauge-inv}, we
proposed a procedure to construct gauge-invariant variables for
higher-order perturbations.
Our starting point to construct gauge-invariant variables is the
following conjecture for the linear-order metric perturbation
$h_{ab}$ defined by Eq.~(\ref{eq:metric-expansion}):
\begin{conjecture}
  \label{conjecture:decomposition-conjecture}
  If there is a symmetric tensor field $h_{ab}$ of the second
  rank, whose gauge transformation rule is
  \begin{eqnarray}
    {}_{{\mathcal Y}}\!h_{ab}
    -
    {}_{{\mathcal X}}\!h_{ab}
    =
    {\pounds}_{\xi_{(1)}}g_{ab},
    \label{eq:linear-metric-gauge-trans}
  \end{eqnarray}
  then there exist a tensor field ${\mathcal H}_{ab}$ and a vector
  field $X^{a}$ such that $h_{ab}$ is decomposed as 
  \begin{eqnarray}
    h_{ab} =: {\mathcal H}_{ab} + {\pounds}_{X}g_{ab},
    \label{eq:linear-metric-decomp}
  \end{eqnarray}
  where ${\mathcal H}_{ab}$ and $X^{a}$ are transformed as
  \begin{equation}
    {}_{{\mathcal Y}}\!{\mathcal H}_{ab} - {}_{{\mathcal X}}\!{\mathcal H}_{ab} =  0, 
    \quad
    {}_{\quad{\mathcal Y}}\!X^{a} - {}_{{\mathcal X}}\!X^{a} = \xi^{a}_{(1)} 
    \label{eq:linear-metric-decomp-gauge-trans}
  \end{equation}
  under the gauge transformation
  (\ref{eq:linear-metric-gauge-trans}), respectively.
\end{conjecture}
In this conjecture, ${\mathcal H}_{ab}$ is gauge-invariant and we
call ${\mathcal H}_{ab}$ as {\it gauge-invariant part} of the
perturbation $h_{ab}$.
On the other hand, the vector field $X^{a}$ in
Eq.~(\ref{eq:linear-metric-decomp}) is gauge dependent, and we
call $X^{a}$ as {\it gauge-variant part} of the perturbation
$h_{ab}$.


The main purpose of this paper is to give an outline of a proof 
of Conjecture \ref{conjecture:decomposition-conjecture}.
In the case of cosmological perturbations on a homogeneous and
isotropic universe, we confirmed that Conjecture
\ref{conjecture:decomposition-conjecture} is almost correct
except for some special modes of perturbations, and then we
could develop the second-order cosmological perturbation theory
in a gauge-invariant manner~\cite{kouchan-cosmo-second}.
On the other hand, in the case of the perturbation theory on an 
arbitrary background spacetime, this conjecture is a highly
non-trivial statement due to the non-trivial curvature of the
background spacetime, though its inverse statement is trivial. 
We will see this situation in detail in
Sec.~\ref{sec:K.Nakamura-2010-3}.


Before going to our outline of a proof of Conjecture
\ref{conjecture:decomposition-conjecture}, we explain how the
higher-order gauge-invariant perturbation theory is developed
based on this conjecture, here.
Through this explanation, we emphasize the importance of
Conjecture \ref{conjecture:decomposition-conjecture}.


As shown in Ref.~\cite{kouchan-gauge-inv}, the second-order 
metric perturbations $l_{ab}$ are decomposed into
gauge-invariant and gauge-variant parts through Conjecture
\ref{conjecture:decomposition-conjecture}.
Actually, using the gauge-variant part $X^{a}$ of the
linear-order metric perturbation $h_{ab}$, we consider the
tensor field $\hat{L}_{ab}$ defined by 
\begin{eqnarray}
  \label{eq:hatLab-def}
  \hat{L}_{ab} := l_{ab} - 2 {\pounds}_{X}h_{ab} + {\pounds}_{X}^{2}g_{ab}.
\end{eqnarray}
Through the gauge-transformation rules (\ref{eq:Bruni-49-one})
and (\ref{eq:linear-metric-decomp-gauge-trans}) for $l_{ab}$ and 
$X^{a}$, respectively, the gauge-transformation rule for this
variable $\hat{L}_{ab}$ is given by 
\begin{eqnarray}
  {}_{{\mathcal Y}}\!\hat{L}_{ab} - {}_{{\mathcal X}}\!\hat{L}_{ab}
  = {\pounds}_{\sigma}g_{ab}, 
  \quad
  \sigma^{a} := \xi^{a}_{(2)} + \left[\xi_{(1)},X\right]^{a}.
\end{eqnarray}
This is identical to the gauge-transformation rule
(\ref{eq:linear-metric-gauge-trans}) in Conjecture
\ref{conjecture:decomposition-conjecture}.
Therefore, we may apply Conjecture
\ref{conjecture:decomposition-conjecture} to the variable
$\hat{L}_{ab}$ and we can decompose it as
\begin{eqnarray}
  \hat{L}_{ab} = {\mathcal L}_{ab} + \pounds_{Y}g_{ab},
  \label{eq:second-hatLab-decomp}
\end{eqnarray}
where the gauge-transformation rules for ${\mathcal L}_{ab}$ and
$Y^{a}$ are given by
\begin{eqnarray}
  \label{eq:second-metric-decomp-gauge-trans}
  {}_{{\mathcal Y}}\!{\mathcal L}_{ab} - {}_{{\mathcal X}}\!{\mathcal L}_{ab} = 0,
  \quad
  {}_{{\mathcal Y}}\!Y^{a} - {}_{{\mathcal X}}\!Y^{a}
  = \xi_{(2)}^{a} + [\xi_{(1)},X]^{a}.
\end{eqnarray}
Thus, we have accomplished the decomposition of the second-order
metric perturbation $l_{ab}$ as
\begin{eqnarray}
  \label{eq:H-ab-in-gauge-X-def-second-1}
  l_{ab}
  =:
  {\mathcal L}_{ab} + 2 {\pounds}_{X} h_{ab}
  + \left(
      {\pounds}_{Y}
    - {\pounds}_{X}^{2} 
  \right)
  g_{ab}.
\end{eqnarray}


Furthermore, as shown in Ref.~\cite{kouchan-gauge-inv}, using
the first- and second-order gauge-variant parts, $X^{a}$ and
$Y^{a}$, of the metric perturbations, the gauge-invariant
variables for an arbitrary tensor field $Q$ other than the
metric are given by 
\begin{eqnarray}
  \label{eq:matter-gauge-inv-def-1.0}
  {}^{(1)}\!{\mathcal Q} &:=& {}^{(1)}\!Q - {\pounds}_{X}Q_{0}
  , \\ 
  \label{eq:matter-gauge-inv-def-2.0}
  {}^{(2)}\!{\mathcal Q} &:=& {}^{(2)}\!Q - 2 {\pounds}_{X} {}^{(1)}Q 
  - \left\{ {\pounds}_{Y} - {\pounds}_{X}^{2} \right\} Q_{0}
  .
\end{eqnarray}
It is straightforward to confirm that the variables
${}^{(1)}\!{\mathcal Q}$ and ${}^{(2)}\!{\mathcal Q}$ defined by
(\ref{eq:matter-gauge-inv-def-1.0}) and
(\ref{eq:matter-gauge-inv-def-2.0}), respectively, are gauge
invariant under the gauge-transformation rules
(\ref{eq:Bruni-47-one}) and (\ref{eq:Bruni-49-one}),
respectively. 
We have to emphasize that not only gauge-invariant parts of the
metric perturbations but also gauge-variant parts $X^{a}$ and
$Y^{a}$ for metric perturbations play crucial role in these
systematic construction of gauge invariant variables
${}^{(1)}\!{\mathcal Q}$ and ${}^{(2)}\!{\mathcal Q}$ through
Eqs.~(\ref{eq:matter-gauge-inv-def-1.0}) and 
(\ref{eq:matter-gauge-inv-def-2.0}).


Equations (\ref{eq:matter-gauge-inv-def-1.0}) and
(\ref{eq:matter-gauge-inv-def-2.0}) have an important 
implication.
To see this, we represent these equations as
\begin{eqnarray}
  \label{eq:matter-gauge-inv-decomp-1.0}
  {}^{(1)}\!Q &=& {}^{(1)}\!{\mathcal Q} + {\pounds}_{X}Q_{0}
  , \\ 
  \label{eq:matter-gauge-inv-decomp-2.0}
  {}^{(2)}\!Q  &=& {}^{(2)}\!{\mathcal Q} + 2 {\pounds}_{X} {}^{(1)}Q 
  + \left\{ {\pounds}_{Y} - {\pounds}_{X}^{2} \right\} Q_{0}
  .
\end{eqnarray}
These equations imply that any perturbation of first and second
order can always be decomposed into their gauge-invariant and
gauge-variant parts as
Eqs.~(\ref{eq:matter-gauge-inv-decomp-1.0}) and
(\ref{eq:matter-gauge-inv-decomp-2.0}), respectively. 
The decomposition formulae
(\ref{eq:matter-gauge-inv-decomp-1.0}) and
(\ref{eq:matter-gauge-inv-decomp-2.0}) are important
consequences in our higher-order gauge-invariant perturbation
theory.
Actually, in Ref.~\cite{kouchan-second}, we have derived the
formulae for the perturbations of the spacetime curvatures and 
showed that all of these are decomposed as 
Eqs.~(\ref{eq:matter-gauge-inv-decomp-1.0}) and
(\ref{eq:matter-gauge-inv-decomp-2.0}).
In addition to the spacetime curvatures, in
Ref.~\cite{kouchan-second-cosmo-matter}, we also summarized the 
formulae for the perturbations of the energy-momentum tensors
for a single scalar field, a perfect fluid, and an imperfect
fluid, and showed that all these energy-momentum tensors and the
equations of motion for the matter fields are decomposed into
their gauge-invariant and gauge-variant parts as 
Eqs.~(\ref{eq:matter-gauge-inv-decomp-1.0}) and
(\ref{eq:matter-gauge-inv-decomp-2.0}). 
As a result of these decompositions, we can easily show that
order by order perturbative equations for any equation on an
arbitrary background spacetime are automatically given in
gauge-invariant
form~\cite{kouchan-second,kouchan-second-cosmo-matter}.
Furthermore, we explicitly derived of the second-order Einstein
equations for cosmological perturbations in a gauge-invariant
manner~\cite{kouchan-cosmo-second,kouchan-second-cosmo-matter}.


We can also expect that the similar structure of equations of
the systems will be maintained in any order perturbations and
our general framework will be applicable to any order
general-relativistic perturbations.
Actually, decomposition formulae for the third-order
perturbations in two-parameter case, which correspond to
Eqs.~(\ref{eq:matter-gauge-inv-decomp-1.0}) and
(\ref{eq:matter-gauge-inv-decomp-2.0}), are explicitly given in
Ref.~\cite{kouchan-gauge-inv}.
Therefore, the similar development is possible for the
third-order perturbations.
Since we could not find any difficulties to extend higher-order
perturbations~\cite{kouchan-gauge-inv} except for the necessity
of long cumbersome calculations, we can construct any order
perturbation theory in gauge-invariant manner, recursively.


We note that, through the above recursive procedure, we can find
gauge-invariant variables for any perturbative variables without
any gauge fixing.
The concept of gauge invariance of perturbations should be
equivalent to ``complete gauge-fixing''.
Therefore, we may say that our procedure gives a complete
gauge-fixing without any explicit gauge-fixing.
We also note that the specification of gauge-variant parts is
not unique.
The different specifications of gauge-variant variables $X^{a}$
and $Y^{a}$ correspond to the different gauge-fixing as
explicitly shown in
Ref.~\cite{Christopherson-Malik-Matravers-Nakamura-comparison}.
In many literature, the explicit gauge-fixing procedures were
proposed and these correspond to the explicit specification of
the gauge-variant parts $X^{a}$ and $Y^{a}$.
However, in this paper, we do not explicitly specify these
gauge-variant variables, though we can specify these
gauge-variant variables at any time.
This is due to the fact that the key idea of our recursive
procedure to find gauge-invariant variables is not in the
explicit form of the gauge-variant parts but in the
gauge-transformation rules of gauge-variant variables of metric
perturbations. 
For example, if we explicitly specify the first-order
gauge-variant part $X^{a}$, it will be difficult to construct
gauge-invariant variables for the second-order metric
perturbations, because the gauge-transformation rule of the
gauge-variant variable $X^{a}$ is used to find gauge-invariant
variables for the second-order metric perturbations in our
recursive procedure.


Finally, we have to emphasize that the above general framework
of the higher-order gauge-invariant perturbation theory are
independent of the explicit form of the background metric
$g_{ab}$ except for Conjecture
\ref{conjecture:decomposition-conjecture}, and are valid not
only in cosmological perturbation case but also the other
generic situations if Conjecture
\ref{conjecture:decomposition-conjecture} is true. 
This implies that if we prove Conjecture
\ref{conjecture:decomposition-conjecture} for an arbitrary
background spacetime in some sense, the above general framework
is applicable to perturbation theories on any background
spacetime.
This is the reason why we proposed Conjecture
\ref{conjecture:decomposition-conjecture} in
Ref.~\cite{kouchan-second-cosmo-matter}.
In the next section, we give a scenario of a proof of Conjecture 
\ref{conjecture:decomposition-conjecture} on an arbitrary
background spacetime which admits ADM decomposition.


\section{Decomposition of the linear-order metric perturbation}
\label{sec:K.Nakamura-2010-3}


Here, we give a scenario of a proof of Conjecture
\ref{conjecture:decomposition-conjecture} in
Sec.~\ref{sec:K.Nakamura-2010-3.1} through the assumption of the
existence of some Green functions of elliptic type derivative
operators.
This scenario is just an extension of the proof in the case of
cosmological perturbations.
The comparison with the case of cosmological perturbations in
Refs.~\cite{kouchan-cosmo-second} is discussed in 
Sec.~\ref{sec:K.Nakamura-2010-3.2}.


\subsection{A scenario of a proof of Conjecture \ref{conjecture:decomposition-conjecture}}
\label{sec:K.Nakamura-2010-3.1}


Now, we give a scenario of a proof of Conjecture
\ref{conjecture:decomposition-conjecture} on an arbitrary
background spacetime.
To do this, we assume that the background spacetimes admit ADM
decomposition.
Therefore, the background spacetime ${\mathcal M}_{0}$ (at least the
portion of ${\mathcal M}_{0}$ which we are addressing) considered
here is $n+1$-dimensional spacetime which is described by the
direct product $\RF\times\Sigma$. 
Here, $\RF$ is a time direction and $\Sigma$ is the spacelike
hypersurface ($\dim\Sigma = n$) embedded in ${\mathcal M}_{0}$. 
This means that ${\mathcal M}_{0}$ is foliated by the one-parameter
family of spacelike hypersurface $\Sigma(t)$, where $t\in\RF$ is
a time function.
In this setup, the metric on ${\mathcal M}_{0}$ is described by
\begin{eqnarray}
  \label{eq:gdb-decomp-dd-minus-main}
  g_{ab} &=& - \alpha^{2} (dt)_{a} (dt)_{b}
  + q_{ij}
  (dx^{i} + \beta^{i}dt)_{a}
  (dx^{j} + \beta^{j}dt)_{b},
\end{eqnarray}
where $\alpha$ is the lapse function, $\beta^{i}$ is the
shift vector, and $q_{ab}=q_{ij}(dx^{i})_{a}(dx^{i})_{b}$ is the
metric on $\Sigma(t)$.


Since the ADM decomposition (\ref{eq:gdb-decomp-dd-minus-main})
of the metric is a local decomposition, we may regard that the
arguments in this paper are restricted to that for a single
patch in ${\mathcal M}_{0}$ which is covered by the metric
(\ref{eq:gdb-decomp-dd-minus-main}). 
Further, we may change the region which is covered by the metric
(\ref{eq:gdb-decomp-dd-minus-main}) through the choice of the
lapse function $\alpha$ and the shift vector $\beta^{i}$.
The choice of $\alpha$ and $\beta^{i}$ is regarded as the
first-kind gauge choice explained in 
Sec.~\ref{sec:Notion_of_gauge_in_general_relativity}, which have
nothing to do with the second-kind gauge as emphasized in
Sec.~\ref{sec:Notion_of_gauge_in_general_relativity}.
Since we may regard that the representation 
(\ref{eq:gdb-decomp-dd-minus-main}) of the background metric is
that on a single patch in ${\mathcal M}_{0}$, in general situation,
each $\Sigma$ may have its boundary $\partial\Sigma$.
For example, in asymptotically flat spacetimes, $\partial\Sigma$
includes asymptotically flat regions~\cite{Wald-book}.
Furthermore, if necessary, we may regard that $\Sigma(t)$ is a
portion of the spacelike hypersurface in ${\mathcal M}_{0}$ and add
disjoint components to the boundary $\partial\Sigma$.
For example, when the formation of black holes occurs, we may
exclude the region inside the black holes from $\Sigma$.
In any case, when we consider the spacelike hypersurface
$\Sigma$ with boundary $\partial\Sigma$, we have to impose
appropriate boundary conditions at the boundary
$\partial\Sigma$.


To consider the decomposition (\ref{eq:linear-metric-decomp}) of
$h_{ab}$, we first consider the components of the metric
$h_{ab}$ as
\begin{eqnarray}
  \label{eq:hab-ADM-decomp}
  h_{ab}
  =
  h_{tt} (dt)_{a}(dt)_{b}
  + 2 h_{ti} (dt)_{(a}(dx^{i})_{b)}
  + h_{ij} (dx^{i})_{a}(dx^{j})_{b}.
\end{eqnarray}
The components $h_{tt}$, $h_{ti}$, and $h_{ij}$ are regarded as
a scalar function, components of a vector field, and the
components of a symmetric tensor field on the spacelike
hypersurface $\Sigma$, respectively.
Under the gauge-transformation rule
(\ref{eq:linear-metric-gauge-trans}), the components $\{h_{tt}$,
$h_{ti}$, $h_{ij}\}$ are transformed as
\begin{eqnarray}
  {}_{{\mathcal Y}}h_{tt}
  -
  {}_{{\mathcal X}}h_{tt}
  &=&
    2 \partial_{t}\xi_{t}
  - \frac{2}{\alpha}\left(
    \partial_{t}\alpha 
    + \beta^{i}D_{i}\alpha 
    - \beta^{j}\beta^{i}K_{ij}
  \right) \xi_{t}
  \nonumber\\
  &&
  - \frac{2}{\alpha} \left(
    \beta^{i}\beta^{k}\beta^{j} K_{kj}
    - \beta^{i} \partial_{t}\alpha
    + \alpha q^{ij} \partial_{t}\beta_{j}
  \right.
  \nonumber\\
  && \quad\quad\quad
  \left.
    + \alpha^{2} D^{i}\alpha 
    - \alpha \beta^{k} D^{i} \beta_{k}
    - \beta^{i} \beta^{j} D_{j}\alpha 
  \right)\xi_{i}
  \label{eq:gauge-trans-of-htt-ADM-BG}
  , \\
  {}_{{\mathcal Y}}h_{ti}
  -
  {}_{{\mathcal X}}h_{ti}
  &=&
  \partial_{t}\xi_{i}
  + D_{i}\xi_{t}
  - \frac{2}{\alpha} \left(
    D_{i}\alpha 
    - \beta^{j}K_{ij}
  \right) \xi_{t}
  - \frac{2}{\alpha} M_{i}^{\;\;j} \xi_{j}
  \label{eq:gauge-trans-of-hti-ADM-BG}
  , \\
  {}_{{\mathcal Y}}h_{ij}
  -
  {}_{{\mathcal X}}h_{ij}
  &=&
  2 D_{(i}\xi_{j)}
  + \frac{2}{\alpha} K_{ij} \xi_{t}
  - \frac{2}{\alpha} \beta^{k} K_{ij} \xi_{k}
  \label{eq:gauge-trans-of-hij-ADM-BG}
  ,
\end{eqnarray}
where $M_{i}^{\;\;j}$ is defined by 
\begin{eqnarray}
  \label{eq:Lij-def}
  M_{i}^{\;\;j}
  :=
  - \alpha^{2} K^{j}_{\;\;i}
  + \beta^{j}\beta^{k} K_{ki}
  - \beta^{j} D_{i}\alpha 
  + \alpha D_{i}\beta^{j}. 
\end{eqnarray}
Here, $K_{ij}$ is the components of the extrinsic curvature of 
$\Sigma$ in ${\mathcal M}_{0}$ and $D_{i}$ is the covariant
derivative associate with the metric $q_{ij}$ ($D_{i}q_{jk}=0$). 
The extrinsic curvature $K_{ij}$ and its trace $K$ are related
to the time derivative of the metric $q_{ij}$ by  
\begin{equation}
  \label{eq:derivative_extrinsic_minus}
  K_{ij} = - \frac{1}{2\alpha}
  \left[\frac{\partial}{\partial t} q_{ij} -
    D_{i}\beta_{j} - D_{j}\beta_{i}\right],
  \quad K := q^{ij}K_{ij}.
\end{equation}


We also note that the gauge-transformation rules
(\ref{eq:gauge-trans-of-htt-ADM-BG})--(\ref{eq:gauge-trans-of-hij-ADM-BG})
represent the gauge-transformation of the second kind which have
nothing to do with the gauge-degree of freedom of the first kind
as explained in
Sec.~\ref{sec:Notion_of_gauge_in_general_relativity}.
We have to emphasize that the main purpose of this paper is to
show how to exclude this gauge degree of freedom of the second
kind inspecting gauge-transformation rules 
(\ref{eq:gauge-trans-of-htt-ADM-BG})--(\ref{eq:gauge-trans-of-hij-ADM-BG}).


The essence of our strategy for the explicit construction of the 
gauge-invariant and gauge-variant parts of the linear metric
perturbation is already given in our short
paper~\cite{kouchan-decomp-letter-version}.
In Ref.~\cite{kouchan-decomp-letter-version}, we consider the
simple case where $\alpha=1$ and $\beta^{i}=0$.
Our strategy is as follows: we first assume that the existence
of the variables $X_{t}$ and $X_{i}$ whose gauge-transformation
rules are given by
\begin{eqnarray}
  &&
  {}_{{\mathcal Y}}X_{t}
  -
  {}_{{\mathcal X}}X_{t}
  =
  \xi_{t}
  \label{eq:K.Nakamura-2010-note-B-14}
  , \\
  &&
  {}_{{\mathcal Y}}X_{i}
  -
  {}_{{\mathcal X}}X_{i}
  =
  \xi_{i}
  \label{eq:K.Nakamura-2010-note-B-15}
  .
\end{eqnarray}
This assumption is confirmed through the explicit construction
of the gauge-variant part of the linear-order metric
perturbation below. 
Similar technique is given by Pereira et
al.~\cite{Pereira:2007yy} in the perturbations on Bianchi type I
cosmology.


Inspecting gauge-transformation rules
(\ref{eq:gauge-trans-of-htt-ADM-BG})--(\ref{eq:gauge-trans-of-hij-ADM-BG}),
we define the symmetric tensor field $\hat{H}_{ab}$ whose
components are given by 
\begin{eqnarray}
  \hat{H}_{tt}
  &:=&
  h_{tt} 
  + \frac{2}{\alpha}\left(
    \partial_{t}\alpha 
    + \beta^{i}D_{i}\alpha 
    - \beta^{j}\beta^{i}K_{ij}
  \right) X_{t}
  \nonumber\\
  &&
  + \frac{2}{\alpha} \left(
    \beta^{i}\beta^{k}\beta^{j} K_{kj}
    - \beta^{i} \partial_{t}\alpha
    + \alpha q^{ij} \partial_{t}\beta_{j}
  \right.
  \nonumber\\
  && \quad\quad\quad
  \left.
    + \alpha^{2} D^{i}\alpha 
    - \alpha \beta^{k} D^{i} \beta_{k}
    - \beta^{i} \beta^{j} D_{j}\alpha 
  \right)X_{i}
  \label{eq:hatHtt-def-generic}
  , \\
  \hat{H}_{ti}
  &:=&
  h_{ti}
  + \frac{2}{\alpha} \left(
    D_{i}\alpha 
    - \beta^{j}K_{ij}
  \right) X_{t}
  + \frac{2}{\alpha} M_{i}^{\;\;j} X_{j}
  \label{eq:hatHti-def-generic}
  , \\
  \hat{H}_{ij}
  &:=&
  h_{ij}
  - \frac{2}{\alpha} K_{ij} X_{t}
  + \frac{2}{\alpha} \beta^{k} K_{ij} X_{k}
  \label{eq:hatHij-def-generic}
  .
\end{eqnarray}
The gauge transformation rules
(\ref{eq:gauge-trans-of-htt-ADM-BG})--(\ref{eq:gauge-trans-of-hij-ADM-BG})
and our assumptions (\ref{eq:K.Nakamura-2010-note-B-14}) and
(\ref{eq:K.Nakamura-2010-note-B-15}) give the
gauge-transformation rules of the components of $\hat{H}_{ab}$
as follows:
\begin{eqnarray}
  {}_{{\mathcal Y}}\!\hat{H}_{tt}
  -
  {}_{{\mathcal X}}\!\hat{H}_{tt}
  &=&
  2 \partial_{t}\xi_{t}
  ,
  \label{eq:gauge-trans-hatHtt}
  \\
  {}_{{\mathcal Y}}\!\hat{H}_{ti}
  -
  {}_{{\mathcal X}}\!\hat{H}_{ti}
  &=&
  \partial_{t}\xi_{i}
  + D_{i}\xi_{t}
  ,
  \label{eq:gauge-trans-hatHti}
  \\
  {}_{{\mathcal Y}}\!\hat{H}_{ij}
  -
  {}_{{\mathcal X}}\!\hat{H}_{ij}
  &=&
  2 D_{(i}\xi_{j)}
  .
  \label{eq:gauge-trans-hatHij}
\end{eqnarray}


Since the components $\hat{H}_{it}$ and $\hat{H}_{ij}$ are
regarded as componets of a vector and a symmetric tensor on
$\Sigma(t)$, respectively, we may apply York's decomposition
reviewed in Appendix 
\ref{sec:S.Deser-1967-J.W.York.Jr-1973-J.W.York.Jr-1974} to
$\hat{H}_{ti}$ and $\hat{H}_{ij}$:
\begin{eqnarray}
  \label{eq:K.Nakamura-2010-2-generic-4-7}
  \hat{H}_{ti} &=:& D_{i}h_{(VL)} + h_{(V)i}, \quad D^{i}h_{(V)i} = 0,
  \\
  \label{eq:K.Nakamura-2010-2-generic-4-8}
  \hat{H}_{ij} &=:& \frac{1}{n} q_{ij} h_{(L)} + h_{(T)ij}, \quad
  q^{ij} h_{(T)ij} = 0,
  \\
  \label{eq:K.Nakamura-2010-2-generic-4-9}
  h_{(T)ij} &=:& \left(Lh_{(TV)}\right)_{ij} + h_{(TT)ij},
  \quad
  D^{i}h_{(TT)ij} = 0,
\end{eqnarray}
where $(Lh_{(TV)})_{ij}$ is defined by [see
Eq.~(\ref{eq:J.W.York.Jr-1973-3-2}) in Appendix
\ref{sec:S.Deser-1967-J.W.York.Jr-1973-J.W.York.Jr-1974}]
\begin{eqnarray}
  \label{eq:LhTVij-def}
  (Lh_{(TV)})_{ij} := D_{i}h_{(TV)j} + D_{j}h_{(TV)i} - \frac{2}{n}q_{ij}D^{l}h_{(TV)l}.
\end{eqnarray}
Equations (\ref{eq:gauge-trans-hatHti}) and
(\ref{eq:gauge-trans-hatHij}) give the gauge-transformation
rules for the variables $h_{(VL)}$, $h_{(V)i}$, $h_{(L)}$,
$h_{(T)ij}$, $h_{(TV)i}$, and $h_{(TT)ij}$.


First, we consider the gauge-transformation rule
(\ref{eq:gauge-trans-hatHti}) in terms of the decomposition
(\ref{eq:K.Nakamura-2010-2-generic-4-7}):
\begin{eqnarray}
  \label{eq:gauge-trans-hatHti-decomp}
  {}_{{\mathcal Y}}\!\hat{H}_{ti}
  -
  {}_{{\mathcal X}}\!\hat{H}_{ti}
  &=&
  D_{i}\left(
    {}_{{\mathcal Y}}\!h_{(VL)}
    -
    {}_{{\mathcal X}}\!h_{(VL)}
  \right)
  + \left(
    {}_{{\mathcal Y}}\!h_{(V)i}
    -
    {}_{{\mathcal X}}\!h_{(V)i}
  \right)
  =
  \partial_{t}\xi_{i}
  + D_{i}\xi_{t}
  .
\end{eqnarray}
Taking the divergence of this gauge-transformation rule and
through the property $D^{i}h_{(V)i}=0$, we obtain 
\begin{eqnarray}
  \Delta\left(
    {}_{{\mathcal Y}}\!h_{(VL)}
    -
    {}_{{\mathcal X}}\!h_{(VL)}
  \right)
  =
  D^{i}\partial_{t}\xi_{i}
  +
  \Delta\xi_{t}
  \label{eq:gauge-trans-div-hatHti-decomp}
  .
\end{eqnarray}
In this paper, we assume that the existence of the Green
function $\Delta^{-1}$ of the Laplacian $\Delta:=D^{i}D_{i}$.
Then, we easily obtain the gauge-transformation rule for
$h_{(VL)}$ as 
\begin{eqnarray}
  {}_{{\mathcal Y}}h_{(VL)} - {}_{{\mathcal X}}h_{(VL)}
  =
  \xi_{t}
  + \Delta^{-1}D^{k}\partial_{t}\xi_{k}
  ,
  \label{eq:K.Nakamura-2010-note-B-26}
\end{eqnarray}
where we have ignored the modes which belong to the kernel of
the derivative operator $\Delta$.
Substituting Eq.~(\ref{eq:K.Nakamura-2010-note-B-26}) into
Eq.~(\ref{eq:gauge-trans-hatHti-decomp}) we obtain the
gauge-transformation rule for the variable $h_{(V)i}$:
\begin{eqnarray}
  {}_{{\mathcal Y}}h_{(V)i} - {}_{{\mathcal X}}h_{(V)i}
  &=&
    \partial_{t}\xi_{i}
  - D_{i}\Delta^{-1}D^{k}\partial_{t}\xi_{k}
  .
  \label{eq:K.Nakamura-2010-note-B-30}
\end{eqnarray}


The gauge-transformation rules for $h_{(L)}$ and $h_{(T)ij}$ are
given from Eq.~(\ref{eq:gauge-trans-hatHij}). 
Since we consider the decomposition
(\ref{eq:K.Nakamura-2010-2-generic-4-8}), the
gauge-transformation rule (\ref{eq:gauge-trans-hatHij}) is given
by  
\begin{eqnarray}
  {}_{{\mathcal Y}}\!\hat{H}_{ij}
  -
  {}_{{\mathcal X}}\!\hat{H}_{ij}
  &=&
  \frac{1}{n} q_{ij} \left(
    {}_{{\mathcal Y}}\!h_{(L)}
    -
    {}_{{\mathcal X}}\!h_{(L)}
  \right)
  + \left(
    {}_{{\mathcal Y}}\!h_{(T)ij}
    -
    {}_{{\mathcal X}}\!h_{(T)ij}
  \right)
  =
  2 D_{(i}\xi_{j)}
  \label{eq:K.Nakamura-2010-note-B-32}
  .
\end{eqnarray}
Taking the trace of Eq.~(\ref{eq:K.Nakamura-2010-note-B-32}), we
obtain 
\begin{eqnarray}
  {}_{{\mathcal Y}}h_{(L)} - {}_{{\mathcal X}}h_{(L)}
  &=&
  2 D^{i}\xi_{i}
  .
  \label{eq:K.Nakamura-2010-note-B-33}
\end{eqnarray}
The traceless part of Eq.~(\ref{eq:K.Nakamura-2010-note-B-32})
is given by 
\begin{eqnarray}
  {}_{{\mathcal Y}}h_{(T)ij} - {}_{{\mathcal X}}h_{(T)ij}
  &=&
  \left(L\xi\right)_{ij}
  .
  \label{eq:K.Nakamura-2010-note-B-34}
\end{eqnarray}
Note that the variable $h_{(T)ij}$ is also decomposed as
Eq.~(\ref{eq:K.Nakamura-2010-2-generic-4-9}) and the 
gauge-transformation rules for the variable $h_{(T)_{ij}}$ is
given by
\begin{eqnarray}
  {}_{{\mathcal Y}}h_{(T)ij}
  -
  {}_{{\mathcal X}}h_{(T)ij}
  &=&
  \left(L\left({}_{{\mathcal Y}}\!h_{(TV)}-{}_{{\mathcal X}}\!h_{(TV)}\right)\right)_{ij}
  + {}_{{\mathcal Y}}h_{(TT)ij} - {}_{{\mathcal X}}h_{(TT)ij}
  \nonumber\\
  &=&
  \left(L\xi\right)_{ij}
  .
  \label{eq:K.Nakamura-2010-note-B-35}
\end{eqnarray}
Taking the divergence of
Eq.~(\ref{eq:K.Nakamura-2010-note-B-35}), we obtain 
\begin{eqnarray}
  {\mathcal D}^{jl}
  \left(
    {}_{{\mathcal Y}}h_{(TV)l} - {}_{{\mathcal X}}h_{(TV)l} - \xi_{l}
  \right)
  =
  0
  ,
  \label{eq:K.Nakamura-2010-note-B-37}
\end{eqnarray}
where the derivative operator ${\mathcal D}^{ij}$ is defined by 
\begin{eqnarray}
  \label{eq:J.W.York.Jr-1973-1-2-kouchan-3-main}
  {\mathcal D}^{ij}
  :=
  q^{ij}\Delta + \left(1 - \frac{2}{n}\right) D^{i}D^{j} +
  R^{ij}.
\end{eqnarray}
Here, $R^{ij}$ is the Ricci curvature on $\Sigma$.
Properties of the derivative operator ${\mathcal D}^{ij}$ are
discussed in Appendix
\ref{sec:S.Deser-1967-J.W.York.Jr-1973-J.W.York.Jr-1974}. 
Here, we assume the existence of the Green function of the
derivative operator ${\mathcal D}^{ij}$ and ignore the modes which 
belong to the kernel of the derivative operator ${\mathcal D}^{ij}$.
Then, we obtain 
\begin{eqnarray}
  \label{eq:K.Nakamura-2010-note-B-38}
  {}_{{\mathcal Y}}h_{(TV)l} - {}_{{\mathcal X}}h_{(TV)l} = \xi_{l}.
\end{eqnarray}
Substituting Eq.~(\ref{eq:K.Nakamura-2010-note-B-38}) into
(\ref{eq:K.Nakamura-2010-note-B-35}), we obtain 
\begin{eqnarray}
  \label{eq:K.Nakamura-2010-note-B-40}
  {}_{{\mathcal Y}}h_{(TT)ij} - {}_{{\mathcal X}}h_{(TT)ij} = 0.
\end{eqnarray}


In summary, we have obtained the gauge-transformation rules for
the variables $\hat{H}_{tt}$, $h_{(VL)}$, $h_{(V)i}$, $h_{(L)}$,
$h_{(TV)i}$, and $h_{(TT)ij}$ as follows:
\begin{eqnarray}
  {}_{{\mathcal Y}}\!\hat{H}_{tt}
  -
  {}_{{\mathcal X}}\!\hat{H}_{tt}
  &=&
  2 \partial_{t}\xi_{t}
  \label{eq:K.Nakamura-2010-note-B-43}
  , \\
  {}_{{\mathcal Y}}h_{(VL)} - {}_{{\mathcal X}}h_{(VL)}
  &=&
  \xi_{t}
  + \Delta^{-1}D^{k}\partial_{t}\xi_{k}
  \label{eq:K.Nakamura-2010-note-B-44}
  , \\
  {}_{{\mathcal Y}}h_{(V)i} - {}_{{\mathcal X}}h_{(V)i}
  &=&
    \partial_{t}\xi_{i}
  - D_{i}\Delta^{-1}D^{k}\partial_{t}\xi_{k}
  \label{eq:K.Nakamura-2010-note-B-45}
  , \\
  {}_{{\mathcal Y}}h_{(L)} - {}_{{\mathcal X}}h_{(L)}
  &=&
  2 D^{i}\xi_{i}
  \label{eq:K.Nakamura-2010-note-B-46}
  , \\
  {}_{{\mathcal Y}}h_{(TV)l} - {}_{{\mathcal X}}h_{(TV)l}
  &=& \xi_{l}
  \label{eq:K.Nakamura-2010-note-B-48}
  ,\\
  \label{eq:K.Nakamura-2010-note-B-50-1}
  {}_{{\mathcal Y}}h_{(TT)ij} - {}_{{\mathcal X}}h_{(TT)ij} &=& 0.
\end{eqnarray}


Since  the gauge transformation rule
(\ref{eq:K.Nakamura-2010-note-B-48}) coincides with the gauge 
transformation rule (\ref{eq:K.Nakamura-2010-note-B-15}) for the
variable $X_{i}$, we may identify the variable $X_{i}$ with
$h_{(TV)i}$: 
\begin{eqnarray}
  \label{eq:K.Nakamura-2010-note-B-51}
  X_{i} := h_{(TV)i}.
\end{eqnarray}
Thus, we have confirmed the existence of the variable
$X_{i}$. 
Next, we show the existence of the variable $X_{t}$ whose
gauge-transformation rule is given by
Eq.~(\ref{eq:K.Nakamura-2010-note-B-14}). 
Inspecting these gauge transformation rules
(\ref{eq:K.Nakamura-2010-note-B-44}) and
(\ref{eq:K.Nakamura-2010-note-B-48}), we find the definition of
$X_{t}$ as 
\begin{eqnarray}
  X_{t}
  &:=&
  h_{(VL)}
  - \Delta^{-1}D^{k}\partial_{t}h_{(TV)k}
  .
  \label{eq:K.Nakamura-2010-note-B-52}
\end{eqnarray}
Actually, the gauge transformation rule for $X_{t}$ defined by
Eq.~(\ref{eq:K.Nakamura-2010-note-B-52}) is given by
Eq.~(\ref{eq:K.Nakamura-2010-note-B-14}). 
This is the desired property for the variable $X_{t}$.
Thus, we have consistently confirmed the existence of the
variables $X_{t}$ and $X_{i}$ which was assumed in the
definitions
(\ref{eq:hatHtt-def-generic})--(\ref{eq:hatHij-def-generic}) of
the components of the tensor field $\hat{H}_{ab}$.
This is the most non-trivial part of the ingredients of this
paper. 
Definitions (\ref{eq:K.Nakamura-2010-note-B-51}) and
(\ref{eq:K.Nakamura-2010-note-B-52}) also imply that we may
start the construction of the gauge-invariant and gauge-variant
variables from the decompositions of the components $h_{ti}$ and
$h_{ij}$ which are obtained by the substitution of
Eqs.~(\ref{eq:K.Nakamura-2010-note-B-52}) and
(\ref{eq:K.Nakamura-2010-note-B-51}) into
Eqs.~(\ref{eq:hatHti-def-generic}) and
(\ref{eq:hatHij-def-generic}) with the decomposition formulae 
(\ref{eq:K.Nakamura-2010-2-generic-4-7})--(\ref{eq:K.Nakamura-2010-2-generic-4-9}).
This approach is discussed in Sec.~\ref{sec:K.Nakamura-2010-4}.


Now, we construct gauge-invariant variables for the linear-order
metric perturbation $h_{ab}$.
First, the gauge-transformation rule
(\ref{eq:K.Nakamura-2010-note-B-50-1}) shows that $h_{(TT)ij}$
is gauge invariant by itself and we define the gauge-invariant
transverse-traceless tensor by
\begin{eqnarray}
  \label{eq:K.Nakamura-2010-note-B-55}
  \chi_{ij} := h_{(TT)ij}.
\end{eqnarray}
The transverse-traceless property of $\chi_{ij}$ is
automatically given by the definition of $h_{(TT)ij}$ in
Eqs.~(\ref{eq:K.Nakamura-2010-2-generic-4-8}) and
(\ref{eq:K.Nakamura-2010-2-generic-4-9}).
Inspecting the gauge-transformation rules
(\ref{eq:K.Nakamura-2010-note-B-45}) and
(\ref{eq:K.Nakamura-2010-note-B-48}), we define a
gauge-invariant vector mode $\nu_{i}$ by 
\begin{eqnarray}
  \nu_{i}
  &:=&
  h_{(V)i}
  - \partial_{t}h_{(TV)i}
  + D_{i}\Delta^{-1}D^{k}\partial_{t}h_{(TV)k}
  .
  \label{eq:K.Nakamura-2010-note-B-56}
\end{eqnarray}
Actually, we can easily confirm that the variable $\nu_{i}$ is
gauge invariant, i.e., ${}_{{\mathcal Y}}\nu_{i}-{}_{{\mathcal X}}\nu_{i}=0$.
Through the divergenceless property of the variable $h_{(V)i}$,
we easily see the property $D^{i}\nu_{i} = 0$.
The gauge-invariant variables for scalar modes are defined as
follows:
First, inspecting gauge-transformation rules
(\ref{eq:K.Nakamura-2010-note-B-14}) and
(\ref{eq:gauge-trans-hatHtt}), we define the scalar variable
$\Phi$ by
\begin{eqnarray}
  - 2 \Phi
  :=
  \hat{H}_{tt}
  - 2 \partial_{t}X_{t}
  =
  \hat{H}_{tt}
  - 2 \partial_{t}h_{(VL)}
  + 2 \partial_{t}\Delta^{-1}D^{k}\partial_{t}h_{(TV)l}
  .
  \label{eq:K.Nakamura-2010-note-B-62}
\end{eqnarray}
Inspecting the gauge-transformation rules
(\ref{eq:K.Nakamura-2010-note-B-46}) and
(\ref{eq:K.Nakamura-2010-note-B-48}), we also define another 
gauge-invariant variable $\Psi$ by
\begin{eqnarray}
  - 2 n \Psi
  :=
  h_{(L)} - 2 D^{i}X_{i}
  =
  h_{(L)} - 2 D^{i}h_{(TV)i}
  .
  \label{eq:K.Nakamura-2010-note-B-64}
\end{eqnarray}
We can easily confirm the gauge invariance of the variables
$\Phi$ and $\Psi$ through gauge-transformation rules
(\ref{eq:K.Nakamura-2010-note-B-14}),
(\ref{eq:gauge-trans-hatHtt}),
(\ref{eq:K.Nakamura-2010-note-B-46}), and
(\ref{eq:K.Nakamura-2010-note-B-48}).
Here, we choose the factor of $\Psi$ in the definition
(\ref{eq:K.Nakamura-2010-note-B-64}) so that we may regard
$\Phi=\Psi$ is Newton's gravitational potential in the
four-dimensional Newtonian limit.


In terms of these gauge-invariant variables and the variables
$X_{t}$ and $X_{i}$, which are defined by
Eqs.~(\ref{eq:K.Nakamura-2010-note-B-52}) and
(\ref{eq:K.Nakamura-2010-note-B-51}), respectively, the original
components $\{h_{tt}$, $h_{ti}$, $h_{ij}\}$ of the metric
perturbation $h_{ab}$ is given by 
\begin{eqnarray}
  h_{tt}
  &=&
  - 2 \Phi
  + 2 \partial_{t}X_{t}
  - \frac{2}{\alpha}\left(
    \partial_{t}\alpha 
    + \beta^{i}D_{i}\alpha 
    - \beta^{j}\beta^{i}K_{ij}
  \right) X_{t}
  \nonumber\\
  && \quad
  - \frac{2}{\alpha} \left(
    \beta^{i}\beta^{k}\beta^{j} K_{kj}
    - \beta^{i} \partial_{t}\alpha
    + \alpha q^{ij} \partial_{t}\beta_{j}
  \right.
  \nonumber\\
  && \quad\quad\quad\quad
  \left.
    + \alpha^{2} D^{i}\alpha 
    - \alpha \beta^{k} D^{i} \beta_{k}
    - \beta^{i} \beta^{j} D_{j}\alpha 
  \right)X_{i}
  \label{eq:K.Nakamura-2010-note-B-73}
  , \\
  h_{ti}
  &=&
  \nu_{i}
  + D_{i}X_{t} 
  + \partial_{t}X_{i}
  - \frac{2}{\alpha} \left(
    D_{i}\alpha 
    - \beta^{j}K_{ij}
  \right) X_{t}
  - \frac{2}{\alpha} M_{i}^{\;\;j} X_{j}
  \label{eq:K.Nakamura-2010-note-B-74}
  , \\
  h_{ij}
  &=&
  - 2 \Psi q_{ij}
  + \chi_{ij}
  + D_{i}X_{j}
  + D_{j}X_{i}
  + \frac{2}{\alpha} K_{ij} X_{t}
  - \frac{2}{\alpha} \beta^{k} K_{ij} X_{k}
  \label{eq:K.Nakamura-2010-note-B-75}
  .
\end{eqnarray}
Equations
(\ref{eq:K.Nakamura-2010-note-B-73})--(\ref{eq:K.Nakamura-2010-note-B-75})
imply that we may identify the components of the gauge-invariant
variables ${\mathcal H}_{ab}$ and the gauge-variant variable
$X_{a}$ so that 
\begin{eqnarray}
  {\mathcal H}_{tt} := - 2 \Phi
  , \quad
  {\mathcal H}_{ti} := \nu_{i}
  , \quad
  {\mathcal H}_{ij} := - 2 \Psi q_{ij} + \chi_{ij}
\end{eqnarray}
and
\begin{eqnarray}
  X_{a} := X_{t}(dt)_{a} + X_{i}(dx^{i})_{a}.
\end{eqnarray}
These identifications lead to the decomposition formula
(\ref{eq:linear-metric-decomp}) for the linear-order metric
perturbation on an arbitrary background spacetime.


We note that, in this outline of a proof, we assumed the
existence of two Green function of the derivative operators
$\Delta:=D^{i}D_{i}$ and ${\mathcal D}^{ij}$ which is defined by
Eq.~(\ref{eq:J.W.York.Jr-1973-1-2-kouchan-3-main}).
In other words, we have ignored the modes which belong to the
kernel of these derivative operators $\Delta$ and ${\mathcal D}^{jl}$.
We call these modes as {\it zero modes}.
To explicitly specify the Green functions for $\Delta$ and
${\mathcal D}^{ij}$, we have to impose boundary conditions at
boundaries $\partial\Sigma$.
Since the operators $\Delta$ and ${\mathcal D}^{ij}$ are elliptic,
the change of the boundary conditions at $\partial\Sigma$ is
adjusted by functions which belong to the kernel of the
operators $\Delta$ and ${\mathcal D}^{ij}$, i.e., zero modes.
Therefore, we may say that the information for the boundary
conditions for the Green functions $\Delta^{-1}$ and
$\left({\mathcal D}^{ij}\right)^{-1}$ is also included in these zero
modes.
To take these modes into account, the different treatments will
be necessary.
We call this issue as {\it zero-mode problem}.


\subsection{Comparison with the FLRW background case}
\label{sec:K.Nakamura-2010-3.2}


Here, we consider the comparison with the case where the
background spacetime ${\mathcal M}_{0}$ is a homogeneous isotropic
universe discussed in Refs.~\cite{kouchan-cosmo-second} to
clarify differences of the above arguments in
Sec.~\ref{sec:K.Nakamura-2010-3.1} from the well-known
formulation of cosmological perturbations. 
The case of a homogeneous isotropic background universe
corresponds to the case $\alpha=1$, $\beta^{i}=0$, and
$K_{ij}=-Hq_{ij}$, where $H=\partial_{t}a/a$ and $a$ is the
scale factor of the universe.


On this background spacetime, in
Refs.~\cite{kouchan-cosmo-second}, we decomposed the components
$h_{ti}$ and $h_{ij}$ of the metric perturbation $h_{ab}$ as
\begin{eqnarray}
  \label{eq:KN2007-Prog-4.2}
  h_{ti}
  &=&
  \tilde{D}_{i}\tilde{h}_{(VL)} + \tilde{h}_{(V)i}
  , \quad
  \tilde{D}^{i}\tilde{h}_{(V)i} = 0
  , \\
  \label{eq:KN2007-Prog-4.3}
  h_{ij}
  &=&
  a^{2} \tilde{h}_{(L)} \gamma_{ij} + a^{2} \tilde{h}_{(T)ij}
  , \quad
  \gamma^{ij}\tilde{h}_{(T)ij}=0, \\
  \label{eq:KN2007-Prog-4.4}
  \tilde{h}_{(T)ij}
  &=&
  \left(
    \tilde{D}_{i}\tilde{D}_{j} - \frac{1}{n} \gamma_{ij} \tilde{\Delta}
  \right) \tilde{h}_{(TL)} 
  + 2 \tilde{D}_{(i}\tilde{h}_{(TV)j)}
  + \tilde{h}_{(TT)ij}
  , \\
  \label{eq:KN2007-Prog-4.5}
  \tilde{D}^{i}\tilde{h}_{(TV)i} &=& 0
  , \quad
  \tilde{D}^{i}\tilde{h}_{(TT)ij} = 0,
\end{eqnarray}
where $q_{ij}=a^{2}\gamma_{ij}$, $\gamma_{ij}$ is the metric
on a maximally symmetric space, $\tilde{D}_{i}$ is the covariant 
derivative associated with the metric $\gamma_{ij}$, and
$\tilde{\Delta}:=\tilde{D}^{i}\tilde{D}_{i}$. 
In the case where $\alpha=1$, $\beta^{i}=0$, and
$K_{ij}=-Hq_{ij}$, the decomposition
(\ref{eq:KN2007-Prog-4.2})--(\ref{eq:KN2007-Prog-4.5}) are
equivalent to the decomposition
(\ref{eq:K.Nakamura-2010-2-generic-4-7})--(\ref{eq:K.Nakamura-2010-2-generic-4-9})
with
Eqs.~(\ref{eq:hatHtt-def-generic})--(\ref{eq:hatHij-def-generic}),
(\ref{eq:K.Nakamura-2010-note-B-51}), and
(\ref{eq:K.Nakamura-2010-note-B-52}) in this paper.
Therefore, one might think that we may also apply the
decomposition
(\ref{eq:KN2007-Prog-4.2})--(\ref{eq:KN2007-Prog-4.5}) even in
the case of an arbitrary background spacetime.
However, in the case of an arbitrary background spacetime, the
decomposition
(\ref{eq:KN2007-Prog-4.2})--(\ref{eq:KN2007-Prog-4.5}) is
ill-defined.
Actually, if we regard that the decomposition
(\ref{eq:KN2007-Prog-4.2})--(\ref{eq:KN2007-Prog-4.5}) is that
for an arbitrary background spacetime, we cannot separate
$\tilde{h}_{(TL)}$ and $\tilde{h}_{(TV)j}$ due to the
non-trivial curvature terms of the background ${\mathcal M}_{0}$ as
pointed out by Deser~\cite{S.Deser-1967}.
These curvature terms come from the commutation relation between
the covariant derivative $D_{i}$ and the derivative operator 
${\mathcal D}^{ij}$.
This is why we apply the decomposition
(\ref{eq:K.Nakamura-2010-2-generic-4-7})--(\ref{eq:K.Nakamura-2010-2-generic-4-9})
with
Eqs.~(\ref{eq:hatHtt-def-generic})--(\ref{eq:hatHij-def-generic}),
(\ref{eq:K.Nakamura-2010-note-B-51}), and
(\ref{eq:K.Nakamura-2010-note-B-52})
instead of
(\ref{eq:KN2007-Prog-4.2})--(\ref{eq:KN2007-Prog-4.5}).


Furthermore, in Refs.~\cite{kouchan-cosmo-second}, we have
assumed the existence of Green functions of the derivative
operators $\tilde{\Delta}$, $\tilde{\Delta}+(n-1)k$, and
$\tilde{\Delta}+nk$ to guarantee the one to one correspondence 
of the set $\{h_{tt}$, $h_{ti}$, $h_{ij}\}$ and $\{\{h_{tt}$,
$h_{(VL)}$, $h_{(L)}$, $h_{(TL)}\}$, $\{h_{(V)i}$,
$h_{(TV)i}\}$, $h_{(TT)ij}\}$, where $k$ is the curvature
constant on the maximally symmetric space. 
The special modes which belong to the kernel of the derivative
operators $\tilde{\Delta}$, $\tilde{\Delta}+(n-1)k$, and
$\tilde{\Delta} + nk$ were not included in the consideration in
Refs.~\cite{kouchan-cosmo-second}.
On the other hand, in this paper, we ignore the modes which
belong to the kernel of the derivative operator $\Delta$ and
${\mathcal D}^{ij}$.
We note that the modes ignored in this paper coincides with the
modes ignored in Ref.~\cite{kouchan-cosmo-second}.
Trivially, the above operator 
$\tilde{\Delta}:=\tilde{D}^{i}\tilde{D}_{i}$ corresponds to the
Laplacian $\Delta$ in this paper. 
In the case of the maximally symmetric $n$-space, the
Riemann curvature and Ricci curvature are given by 
\begin{eqnarray}
  \label{eq:curvature-tensors-in-maximally-symmetric-n-space}
  R_{ijkl} = 2 k q_{k[i}q_{j]l}
  = 2 k q_{k[i}q_{j]l},
  \quad
  R_{ik} = q^{jl}R_{ijkl}
  = (n-1) k q_{ik}.
\end{eqnarray}
In this case, the derivative operator ${\mathcal D}^{ij}$ defined by
Eq.~(\ref{eq:J.W.York.Jr-1973-1-2-kouchan-3-main}) is given by 
\begin{eqnarray}
  {\mathcal D}^{ij}
  =
  q^{ij}\left(
    \Delta + (n - 1) k
  \right) + \left(1 - \frac{2}{n}\right) D^{i}D^{j}.
\end{eqnarray}
When the operator ${\mathcal D}^{ij}$ acts on an arbitrary
transverse vector field $v_{i}$ ($D^{i}v_{i}=0$), we easily see
that 
\begin{eqnarray}
  {\mathcal D}^{ij}v_{j} = \left( \Delta + (n - 1) k \right) v^{i}.
\end{eqnarray}
Therefore, the kernel of the derivative operator 
$\tilde{\Delta}+(n-1)k$ in Refs.~\cite{kouchan-cosmo-second}
is included in the kernel of the derivative operator 
${\mathcal D}^{ij}$ in this paper. 
Finally, we consider the case where the derivative operator
${\mathcal D}^{jl}$ acts on the gradient $D_{l}f$ of an arbitrary
scalar function $f$:
\begin{eqnarray}
  {\mathcal D}^{jl} D_{l}f
  &=&
  2 \frac{n-1}{n} \left[
      D^{j}\Delta
    + \frac{n}{n-1} R^{jl} D_{l}
  \right] f
  \label{eq:K.Nakamura-2010-note-B-92}
  .
\end{eqnarray}
In the case of maximally symmetric $n$-space, curvature tensors
are given by
Eqs.~(\ref{eq:curvature-tensors-in-maximally-symmetric-n-space})
and the derivative operator ${\mathcal D}^{jl} D_{l}$ is
given by 
\begin{eqnarray}
  {\mathcal D}^{jl} D_{l}f
  &=&
  2 \frac{n-1}{n} D^{j} \left(
    \Delta + n k 
  \right) f
  \label{eq:K.Nakamura-2010-note-B-95}
  .
\end{eqnarray}
To solve the equation ${\mathcal D}^{jl} D_{l}f = g^{j}$, we have to
use the Green functions associated with the derivative operators
$\Delta$ and $\Delta + n k$.
These are the reason for the fact that the Green functions
$\Delta^{-1}$, $\left(\Delta+(n-1)k\right)^{-1}$, and
$\left(\Delta+nk\right)^{-1}$ were necessary to guarantee the 
one-to-one correspondence between the components $\{h_{ti}$,
$h_{ij}\}$ and $\{\tilde{h}_{(VL)}$, $\tilde{h}_{(V)i}$,
$\tilde{h}_{(L)}$, $\tilde{h}_{(TL)}$, $\tilde{h}_{(TV)i}$,
$\tilde{h}_{(TT)ij}\}$ in
Eqs.~(\ref{eq:KN2007-Prog-4.2})--(\ref{eq:KN2007-Prog-4.5}).
In other words, we may say that the special modes belong to the
kernel of the derivative operators $\Delta$ and ${\mathcal D}^{ij}$
are equivalent to the special modes which belong to the kernel
of the derivative operators $\Delta$, $\Delta+(n-1)k$, and
$\Delta+nk$ in the case of the maximally symmetric space
$\Sigma$ in Refs.~\cite{kouchan-cosmo-second}.


Finally, we note that the gauge-invariant variables defined by 
Eqs.~(\ref{eq:K.Nakamura-2010-note-B-55})--(\ref{eq:K.Nakamura-2010-note-B-64}) 
are generalizations of the metric perturbation in the
longitudinal gauge in cosmological perturbation
theory~\cite{Christopherson-Malik-Matravers-Nakamura-comparison}.
As noted in
Refs.~\cite{Nakamura:2010yg,kouchan-second-cosmo-matter,Christopherson-Malik-Matravers-Nakamura-comparison},
the choice of the gauge-invariant variable is not unique and
there are many choice of the gauge-invariant variables.
This situation corresponds to the fact that there are infinitely
many complete gauge-fixing procedures.
Actually, in
Ref.~\cite{Christopherson-Malik-Matravers-Nakamura-comparison},
through the specification of the gauge-variant parts $X^{a}$ and
$Y^{a}$ of the first- and the second-order metric perturbations
in Eqs.~(\ref{eq:linear-metric-decomp}) and
(\ref{eq:H-ab-in-gauge-X-def-second-1}), 
we realized the two different complete gauge-fixing in the
first- and the second-order perturbations, namely, the Poission
gauge and the flat gauge, at kinematical level.
Through the similar technique, Uggla and
Wainwright~\cite{C.Uggla-J.Wainwright-2011} derived the
linear-order Einstein equations both in the longitudinal gauge
(Poisson gauge) and in the flat gauge (constant curvature gauge)
in a compact manner.


\section{Alternative construction of gauge-variant and
  gauge-invariant parts}
\label{sec:K.Nakamura-2010-4}


The result obtained in Sec.~\ref{sec:K.Nakamura-2010-3.1}
implies that we may define the variables $h_{(VL)}$, $h_{(V)i}$,
$h_{(L)}$, $h_{(TV)i}$, $h_{(TT)ij}$ by the following
decomposition formulae for the components $h_{ti}$ and $h_{ij}$:
\begin{eqnarray}
  &&
  h_{ti}
  =:
  D_{i}h_{(VL)} + h_{(V)i}
  \nonumber\\
  && \quad\quad\quad
  - \frac{2}{\alpha} \left(
    D_{i}\alpha 
    - \beta^{k}K_{ik}
  \right) \left(
    h_{(VL)}
    - \Delta^{-1}D^{k}\partial_{t}h_{(TV)k}
  \right)
  - \frac{2}{\alpha} M_{i}^{\;\;k} h_{(TV)k}
  \label{eq:hti-decomp-alternative}
  , \\
  &&
  h_{ij}
  =:
  \frac{1}{n} q_{ij} h_{(L)} + \left(Lh_{(TV)}\right)_{ij} + h_{(TT)ij}
  \nonumber\\
  && \quad\quad\quad
  + \frac{2}{\alpha} K_{ij} \left(
    h_{(VL)}
    - \Delta^{-1}D^{k}\partial_{t}h_{(TV)k}
  \right)
  - \frac{2}{\alpha} K_{ij} \beta^{k} h_{(TV)k}
  \label{eq:hij-decomp-alternative}
  , \\
  && D^{i}h_{(V)i} = 0, \quad q^{ij} h_{(TT)ij} = 0 = D^{i}h_{(TT)ij}
  \label{eq:hti-hij-decomp-conditions}
  ,
\end{eqnarray}
where $M_{i}^{\;\;j}$ is defined by Eq.~(\ref{eq:Lij-def}).
Here, these expressions are obtained through the substitution of 
Eqs.~(\ref{eq:K.Nakamura-2010-note-B-51}) and
(\ref{eq:K.Nakamura-2010-note-B-52}) into
(\ref{eq:hatHti-def-generic}) and (\ref{eq:hatHij-def-generic})
and York's decomposition
(\ref{eq:K.Nakamura-2010-2-generic-4-7})--(\ref{eq:K.Nakamura-2010-2-generic-4-9}).
In this section, we check the consistency of our result in
Sec.~\ref{sec:K.Nakamura-2010-3.1}.
To do this, we change the starting point of our arguments to the
decomposition formulae
(\ref{eq:hti-decomp-alternative})--(\ref{eq:hti-hij-decomp-conditions}),
though the starting point of our outline of a proof in
Sec.~\ref{sec:K.Nakamura-2010-3.1} was
Eqs.~(\ref{eq:hatHtt-def-generic})--(\ref{eq:hatHij-def-generic})
with the assumption of the existence of the variables $X_{t}$
and $X_{i}$.
First, we consider the derivation of the inverse relation of
Eqs.~(\ref{eq:hti-decomp-alternative})--(\ref{eq:hti-hij-decomp-conditions})
in Sec.~\ref{sec:K.Nakamura-2010-4.1}.
Then, we derive the gauge transformation rules for the variables
$h_{(VL)}$, $h_{(V)i}$, $h_{(L)}$, $h_{(TV)i}$, and $h_{(TT)ij}$
in Sec.~\ref{sec:K.Nakamura-2010-4.2}.


\subsection{Inverse relation}
\label{sec:K.Nakamura-2010-4.1}


To derive the inverse relation of
Eqs.~(\ref{eq:hti-decomp-alternative})--(\ref{eq:hti-hij-decomp-conditions}),
we first consider Eq.~(\ref{eq:hti-decomp-alternative}).
Assuming the existence of the Green function ${\mathcal F}^{-1}$
for the elliptic derivative operator 
\begin{eqnarray}
  \label{eq:calF-operator-def}
  {\mathcal F}
  := 
  \Delta
  - \frac{2}{\alpha} \left(
    D_{i}\alpha 
    - \beta^{j}K_{ij}
  \right)D^{i}
  - 2 D^{i}\left\{
    \frac{1}{\alpha} \left(
      D_{i}\alpha 
      - \beta^{j}K_{ij}
    \right)
  \right\},
\end{eqnarray}
we obtain the relations
\begin{eqnarray}
  h_{(VL)}
  &=&
  {\mathcal F}^{-1}\left[
      D^{k}h_{tk}
    - D^{k}\partial_{t}h_{(TV)k}
    + D^{k}\left(
      \frac{2}{\alpha} M_{k}^{\;\;l} h_{(TV)l}
    \right)
  \right]
  \nonumber\\
  &&
  + \Delta^{-1}D^{k}\partial_{t}h_{(TV)k}
  ,
  \label{eq:K.Nakamura-2010-note-IV-83}
  \\
  h_{(V)i}
  &=&
  h_{ti}
  - D_{i}\Delta^{-1}D^{k}\partial_{t}h_{(TV)k}
  + \frac{2}{\alpha} M_{i}^{\;\;k} h_{(TV)k}
  \nonumber\\
  &&
  + \left[
    D_{i}
    - \frac{2}{\alpha} \left(
      D_{i}\alpha 
      - \beta^{j}K_{ij}
    \right)
  \right]
  {\mathcal F}^{-1}\left[\frac{}{}
    - D^{k}h_{tk}
    + D^{k}\partial_{t}h_{(TV)k}
  \right.
  \nonumber\\
  && \quad\quad\quad\quad\quad\quad\quad\quad\quad\quad\quad\quad\quad\quad\quad
  \left.
    - D^{k}\left(
      \frac{2}{\alpha} M_{k}^{\;\;l} h_{(TV)l}
    \right)
  \right]
  .
  \label{eq:K.Nakamura-2010-note-IV-85}
\end{eqnarray}
Equations (\ref{eq:K.Nakamura-2010-note-IV-83}) and
(\ref{eq:K.Nakamura-2010-note-IV-85}) imply that we can obtain
the relations between $\{h_{(VL)},h_{(V)i}\}$ and $\{h_{ti}$,
$h_{ij}\}$ if the relation between $h_{(TV)i}$ and $\{h_{ti}$,
$h_{ij}\}$ is specified.
On the other hand, the trace part and the traceless part of
Eq.~(\ref{eq:hij-decomp-alternative}) are given by 
\begin{eqnarray}
  &&
  h_{(L)} 
  =
  q^{ij}h_{ij}
  + \frac{2}{\alpha} K \beta^{k} h_{(TV)k}
  \nonumber\\
  && \quad\quad\quad
  - \frac{2}{\alpha} K \left(
    {\mathcal F}^{-1}\left[
      D^{k}h_{tk}
      - D^{k}\partial_{t}h_{(TV)k}
      + D^{k}\left(
        \frac{2}{\alpha} M_{k}^{\;\;l} h_{(TV)l}
      \right)
    \right]
  \right)
  ,
  \label{eq:K.Nakamura-2010-note-IV-87}
  \\
  &&
  h_{ij}
  - \frac{1}{n} q_{ij} q^{kl}h_{kl}
  =
  \left(Lh_{(TV)}\right)_{ij}
  + h_{(TT)ij}
  - \frac{2}{\alpha} \tilde{K}_{ij} \beta^{k} h_{(TV)k}
  \nonumber\\
  && \quad\quad\quad\quad\quad\quad\quad\quad
  + \frac{2}{\alpha} \tilde{K}_{ij} {\mathcal F}^{-1}\left[ \frac{}{}
    D^{k}h_{tk}
    - D^{k}\partial_{t}h_{(TV)k}
  \right.
  \nonumber\\
  && \quad\quad\quad\quad\quad\quad\quad\quad\quad\quad\quad\quad\quad\quad
  \left.
    + D^{k}\left(
      \frac{2}{\alpha} M_{k}^{\;\;l} h_{(TV)l}
    \right)
  \right]
  ,
  \label{eq:K.Nakamura-2010-note-IV-88}
\end{eqnarray}
where we have used Eq.~(\ref{eq:K.Nakamura-2010-note-IV-83}) and
defined the traceless part $\tilde{K}_{ij}$ of the extrinsic
curvature $K_{ij}$ by 
\begin{eqnarray}
  \label{eq:K.Nakamura-2010-note-IV-89}
  \tilde{K}_{ij} := K_{ij} - \frac{1}{n} q_{ij} K.
\end{eqnarray}
Taking the divergence of
Eq.~(\ref{eq:K.Nakamura-2010-note-IV-88}), we obtain the single
integro-differential equation for $h_{(TV)k}$:
\begin{eqnarray}
  &&
  {\mathcal D}_{j}^{\;\;k}h_{(TV)k}
  + D^{m}\left[
    \frac{2}{\alpha} \tilde{K}_{mj} \left\{ \frac{}{}
      {\mathcal F}^{-1}D^{k}\left(
        \frac{2}{\alpha} M_{k}^{\;\;l} h_{(TV)l}
        - \partial_{t}h_{(TV)k}
      \right)
    \right.
  \right.
  \nonumber\\
  && \quad\quad\quad\quad\quad\quad\quad\quad\quad\quad\quad\quad
  \left.
    \left.
      - \beta^{k} h_{(TV)k}
      \frac{}{}
    \right\}
  \right]
  \nonumber\\
  &=&
  D^{m}\left[
    h_{mj}
    - \frac{1}{n} q_{mj} q^{kl}h_{kl}
    - \frac{2}{\alpha} \tilde{K}_{mj} {\mathcal F}^{-1}D^{k}h_{tk}
  \right]
  .
  \label{eq:K.Nakamura-2010-note-IV-92}
\end{eqnarray}


The existence and the uniqueness of the solution to this
integro-differential equation is highly non-trivial.
However, we assume the existence and the uniqueness of the
solution $h_{(TV)k}=h_{(TV)k}\left[h_{tm},h_{mn}\right]$ to this 
integro-differential equation 
(\ref{eq:K.Nakamura-2010-note-IV-92}), here.
This solution describes the expression of the variable
$h_{(TV)i}$ in terms of the original components $h_{ti}$ and
$h_{ij}$ of the metric perturbation $h_{ab}$.
Substituting the solution
$h_{(TV)k}=h_{(TV)k}\left[h_{tm},h_{mn}\right]$ to
Eq.~(\ref{eq:K.Nakamura-2010-note-IV-92}) into
Eqs.~(\ref{eq:K.Nakamura-2010-note-IV-83})--(\ref{eq:K.Nakamura-2010-note-IV-87}),
we can obtain the representation of the variables $h_{(VL)}$,
$h_{(V)i}$, $h_{(L)}$ in terms of the original components
$h_{ti}$ and $h_{ij}$ of $h_{ab}$.
Furthermore, the representation of the variable $h_{(TT)ij}$ in
terms of $h_{ti}$ and $h_{ij}$ are derived from
Eq.~(\ref{eq:K.Nakamura-2010-note-IV-88}) through the
substitution of the solution
$h_{(TV)k}=h_{(TV)k}\left[h_{tm},h_{mn}\right]$ to
Eq.~(\ref{eq:K.Nakamura-2010-note-IV-92}).


Thus, the decomposition formulae
(\ref{eq:hti-decomp-alternative})--(\ref{eq:hti-hij-decomp-conditions})
are invertible if the Green functions $\Delta^{-1}$, 
${\mathcal F}^{-1}$ exist and the solution to the
integro-differential equation
(\ref{eq:K.Nakamura-2010-note-IV-92}) exists and is unique.


\subsection{Gauge-transformation rules}
\label{sec:K.Nakamura-2010-4.2}


Through similar calculations to those in
Sec.~\ref{sec:K.Nakamura-2010-4.1}, we can derive the
gauge-transformation rules for the variables $h_{(VL)}$, 
$h_{(V)i}$, $h_{(L)}$, $h_{(TV)i}$, and $h_{(TT)ij}$.
From Eqs.~(\ref{eq:K.Nakamura-2010-note-IV-83}) and
(\ref{eq:K.Nakamura-2010-note-IV-85}) the gauge-transformation
rules (\ref{eq:gauge-trans-of-hti-ADM-BG}) for the component
$h_{it}$, we obtain the gauge-transformation rule for the
variable $h_{(VL)}$ and $h_{(V)i}$:
\begin{eqnarray}
  {}_{{\mathcal Y}}h_{(VL)}
  -
  {}_{{\mathcal X}}h_{(VL)}
  &=&
    \xi_{t}
  + \Delta^{-1}D^{k}\partial_{t}\xi_{k}
  \nonumber\\
  &&
  + {\mathcal F}^{-1}D^{k}\left[
    - \partial_{t}A_{k}
    + \frac{2}{\alpha} M_{k}^{\;\;l} A_{l}
  \right]
  + \Delta^{-1}D^{k}\partial_{t}A_{k}
  ,
  \label{eq:K.Nakamura-2010-note-IV-101}
  \\
  {}_{{\mathcal Y}}h_{(V)i}
  -
  {}_{{\mathcal X}}h_{(V)i}
  &=&
  \partial_{t}\xi_{i}
  - D_{i}\Delta^{-1}D^{k}\partial_{t}\xi_{k}
  \nonumber\\
  &&
  + \left[
    D_{i}
    - \frac{2}{\alpha} \left(
      D_{i}\alpha 
      - \beta^{j}K_{ij}
    \right)
  \right]
  {\mathcal F}^{-1}D^{k}\left[
     \partial_{t}A_{k}
    - \frac{2}{\alpha} M_{k}^{\;\;l} A_{l}
  \right]
  \nonumber\\
  &&
  - D_{i}\Delta^{-1}D^{k}\partial_{t}A_{k}
  + \frac{2}{\alpha} M_{i}^{\;\;k} A_{k}
  , 
  \label{eq:K.Nakamura-2010-note-IV-103}
\end{eqnarray}
where $A_{i}:={}_{{\mathcal Y}}h_{(TV)i}-{}_{{\mathcal X}}h_{(TV)i}-\xi_{i}$. 
As in the case of the relations
(\ref{eq:K.Nakamura-2010-note-IV-83}) and
(\ref{eq:K.Nakamura-2010-note-IV-85}), these
gauge-transformation rules
(\ref{eq:K.Nakamura-2010-note-IV-101}) and
(\ref{eq:K.Nakamura-2010-note-IV-103}) imply that we can obtain
the gauge-transformation rules for the variable $h_{(VL)}$ and 
$h_{(V)i}$ if the gauge-transformation rule for the variable
$h_{(TV)i}$ is specified.


From Eq.~(\ref{eq:K.Nakamura-2010-note-IV-87}) and the
gauge-transformation rule (\ref{eq:gauge-trans-of-hij-ADM-BG}),
we can derive the gauge-transformation rule for the variable
$h_{(L)}$: 
\begin{eqnarray}
  {}_{{\mathcal Y}}h_{(L)} 
  -
  {}_{{\mathcal X}}h_{(L)} 
  &=&
  2 D^{l}\xi_{l}
  + \frac{2}{\alpha} K \beta^{k} A_{k}
  \nonumber\\
  &&
  + \frac{2}{\alpha} K \left(
    {\mathcal F}^{-1}D^{k}\left[
      \partial_{t}A_{k}
      - \frac{2}{\alpha} M_{k}^{\;\;l} A_{l}
    \right]
  \right)
  .
  \label{eq:K.Nakamura-2010-note-IV-109}
\end{eqnarray}
As in the case of the gauge-transformation rules
(\ref{eq:K.Nakamura-2010-note-IV-101}) and
(\ref{eq:K.Nakamura-2010-note-IV-103}), the gauge-transformation
rule (\ref{eq:K.Nakamura-2010-note-IV-109}) also implies that we
can obtain the gauge-transformation rule for the variable
$h_{(L)}$ if the gauge-transformation rule for the variable 
$h_{(TV)i}$ is specified.
On the other hand, from the gauge-transformation rule for the
traceless part (\ref{eq:K.Nakamura-2010-note-IV-88}) of
$h_{ij}$, we obtain the equation 
\begin{eqnarray}
  &&
  \left(LA\right)_{ij}
  + {}_{{\mathcal Y}}h_{(TT)ij}
  - {}_{{\mathcal X}}h_{(TT)ij}
  - \frac{2}{\alpha} \tilde{K}_{ij} \beta^{k} A_{k}
  \nonumber\\
  && \quad\quad
  - \frac{2}{\alpha} \tilde{K}_{ij} {\mathcal F}^{-1}D^{k}\left[ \frac{}{}
    \partial_{t}A_{k}
    - \frac{2}{\alpha} M_{k}^{\;\;l} A_{k}
  \right]
  = 0
  ,
  \label{eq:K.Nakamura-2010-note-IV-113}
\end{eqnarray}
where we have used Eqs.~(\ref{eq:gauge-trans-of-hti-ADM-BG}) and
(\ref{eq:gauge-trans-of-hij-ADM-BG}). 
The divergence of Eq.~(\ref{eq:K.Nakamura-2010-note-IV-113})
yields
\begin{eqnarray}
  &&
  {\mathcal D}_{j}^{\;\;l}A_{l}
  + D^{l}\left[
    \frac{2}{\alpha} \tilde{K}_{lj} \left\{
      {\mathcal F}^{-1}D^{k}\left(
        \partial_{t}A_{k}
        - \frac{2}{\alpha} M_{k}^{\;\;l} A_{k}
      \right)
      - \beta^{k} A_{k}
    \right\}
  \right]
  = 0
  .
  \label{eq:K.Nakamura-2010-note-IV-114}
\end{eqnarray}


Here, we note that we have assumed the existence and the
uniqueness of the solution to
Eq.~(\ref{eq:K.Nakamura-2010-note-IV-92}).
Since Eq.~(\ref{eq:K.Nakamura-2010-note-IV-114}) is the
homogeneous version of
Eq.~(\ref{eq:K.Nakamura-2010-note-IV-92}), this assumption
yields that we have the unique solution $A_{k}=0$ to
Eq.~(\ref{eq:K.Nakamura-2010-note-IV-114}), i.e.,
\begin{eqnarray}
  \label{eq:K.Nakamura-2010-note-IV-116}
  {}_{{\mathcal Y}}h_{(TV)i} - {}_{{\mathcal X}}h_{(TV)i} = \xi_{i}.
\end{eqnarray}
Thus, we have specified the gauge-transformation rule for the
variable $h_{(TV)i}$ and the gauge-transformation rule
(\ref{eq:K.Nakamura-2010-note-IV-116}) coincides with
Eq.~(\ref{eq:K.Nakamura-2010-note-B-38}).


Substituting Eq.~(\ref{eq:K.Nakamura-2010-note-IV-116}) into
Eqs.~(\ref{eq:K.Nakamura-2010-note-IV-101})--(\ref{eq:K.Nakamura-2010-note-IV-113}),
we obtain the gauge-transformation rules for the variables
$h_{(VL)}$, $h_{(V)i}$, $h_{(L)}$, and $h_{(TT)ij}$.
We easily see that the resulting gauge-transformation rules for
these variables are given by
Eqs.~(\ref{eq:K.Nakamura-2010-note-B-26}),
(\ref{eq:K.Nakamura-2010-note-B-30}),
(\ref{eq:K.Nakamura-2010-note-B-33}), and
(\ref{eq:K.Nakamura-2010-note-B-40}), respectively.
Further, we can construct gauge-variant and gauge-invariant
parts of the metric perturbation $h_{ab}$ in the same manner as
in Sec.~\ref{sec:K.Nakamura-2010-3.1} and we can confirm
Conjecture \ref{conjecture:decomposition-conjecture}.
Thus, we have reached to the same conclusions as those obtained
in Sec.~\ref{sec:K.Nakamura-2010-3.1}.
Therefore, we may say that the results obtained in
Sec.~\ref{sec:K.Nakamura-2010-3.1} are consistent.


\section{Summary and discussions}
\label{sec:summary}


In summary, we proposed a scenario of a proof of Conjecture
\ref{conjecture:decomposition-conjecture} for an arbitrary
background spacetime which admits ADM decomposition.
Conjecture \ref{conjecture:decomposition-conjecture} states that
we already know the procedure to decompose the linear-order
metric perturbation $h_{ab}$ into its gauge-invariant part 
${\mathcal H}_{ab}$ and gauge-variant part $X_{a}$.
In the cosmological perturbation case, this conjecture was
confirmed and then the second-order cosmological perturbation
theory was developed in our series of
papers~\cite{kouchan-cosmo-second,kouchan-second-cosmo-matter,kouchan-second-cosmo-consistency}. 
However, as reviewed in
Sec.~\ref{sec:General-framework-of-the-gauge-invariant-perturbation-theory},
Conjecture \ref{conjecture:decomposition-conjecture} is the only 
non-trivial part when we consider the general framework of
gauge-invariant perturbation theory on an arbitrary background
spacetime. 
Although there will be many approaches to prove Conjecture
\ref{conjecture:decomposition-conjecture}, in this paper, we
just proposed an outline a proof for an arbitrary background
spacetime.


In the outline shown in Sec.~\ref{sec:K.Nakamura-2010-3.1}, we
assumed the existence of Green functions of the elliptic
derivative operators $\Delta$ and ${\mathcal D}^{ij}$.
This assumption implies that we have ignored the modes which
belong to the kernel of these derivative operators.
We call these modes as {\it zero modes}.
To derive the explicit representation of these Green functions,
we have to impose appropriate boundary conditions for the
perturbative metric at the boundary $\partial\Sigma$.
The modes which belong to the kernel of the above derivative
operators also includes the degree of freedom of these boundary
conditions, because of the ellipticity of the derivative
operators $\Delta$ and ${\mathcal D}^{ij}$ as noted in 
Sec.~\ref{sec:K.Nakamura-2010-3.1}.
For this reason, we also emphasized the importance of these zero
modes from a view point of the globalization of gauge-invariant
variables~\cite{K.Nakamura-IJMPD-2012}. 
Within the arguments in this paper, there is no information for
the treatment of these mode.
To discuss these modes, different treatments of perturbations
will be necessary.
We call this issue as {\it zero-mode problem}.
The situation is similar to the cosmological perturbation case
as noted in Sec.~\ref{sec:K.Nakamura-2010-3.2} and zero-mode problem
exists even in the cosmological perturbation case.


This zero-mode problem in cosmological perturbations also
corresponds to the $l=0$ and $l=1$ (even) mode problem in
perturbation theories on spherically symmetric background
spacetimes.
In the perturbation theory on spherically symmetric background
spacetimes, one usually considers $2+2$
formulation~\cite{Gerlach_Sengupta-1979} (or $2+n$
formulation~\cite{Kodama-Ishibashi-2004}), in which the similar
decomposition to
Eqs.~(\ref{eq:KN2007-Prog-4.3})--(\ref{eq:KN2007-Prog-4.5}) is
applied.
In this case, the indices $i,j,...$ in these equations
correspond to the indices of the components of a tensor field on
$S^{2}$. 
Since $S^{2}$ is a 2-dimensional maximally symmetric space with
the positive curvature, we may regard $n=2$ and $k=1$ in
Sec.~\ref{sec:K.Nakamura-2010-3.2}. 
Then, the three derivative operators $\Delta$, $\Delta+(n-1)k$,
and $\Delta+nk$ in Sec.~\ref{sec:K.Nakamura-2010-3.2} are given
by $\Delta$, $\Delta + 1$, and $\Delta + 2$, respectively. 
Since the eigenvalue of the Laplacian $\Delta$ on $S^{2}$ is
given by $\Delta = -l(l+1)$, we may say that the modes with
$l=0$ and $l=1$ belong to the kernel of the derivative operator
$\Delta$ and $\Delta + n k$, respectively.
Therefore, we may say that the problem concerning about the
modes with $l=0$ and $l=1$ in the perturbations on spherically
symmetric background spacetimes is the same problem as the
zero-mode problem mentioned above.


Thus, the arguments in this paper show that zero-mode problem
generally appears in many perturbation theories in general
relativity and we have seen that the appearance of this
zero-mode problem from general point of view.
Furthermore, as discussed in
Appendix~\ref{sec:S.Deser-1967-J.W.York.Jr-1973-J.W.York.Jr-1974},
the kernel of the derivative operator ${\mathcal D}^{ij}$
includes conformal Killing (or Killing) vectors on $\Sigma$.
Therefore, the existence of zero mode is also related to the
symmetry of the background spacetime.
To resolve this zero-mode problem, careful discussions on
domains of functions for perturbations and its boundary
conditions at $\partial\Sigma$ will be necessary. 
We leave this zero-mode problem as a future work.


In our outline of a proof in Sec.~\ref{sec:K.Nakamura-2010-3.1},
we used a tricky logic to construct gauge-invariant and
gauge-variant parts of the metric perturbation $h_{ab}$.
Therefore, in Sec.~\ref{sec:K.Nakamura-2010-4}, we confirmed the
consistency of our result in Sec.~\ref{sec:K.Nakamura-2010-3.1}
through the change of the starting point of our arguments, and
then, we have reached to the same conclusion as in
Sec.~\ref{sec:K.Nakamura-2010-3.1}. 
Due to this fact, we may say that the result obtained in
Sec.~\ref{sec:K.Nakamura-2010-3.1} is consistent.
In the approach in Sec.~\ref{sec:K.Nakamura-2010-4}, we assume
the existence of Green functions $\Delta^{-1}$ and 
${\mathcal F}^{-1}$. 
Further, we also assume the existence and the uniqueness of the 
solution to the integro-differential equation
(\ref{eq:K.Nakamura-2010-note-IV-92}).
Although the correspondence between the existence of the Green
functions which are assumed in
Sec.~\ref{sec:K.Nakamura-2010-3.1} and these assumptions is not
clear, we may say that the above zero-mode problem is delicate
in the case of an arbitrary background spacetime.


We also emphasize that the gauge issue discussed in this paper
is just a kinematical one and we did not use any information of
the field equations such as Einstein equations.
In other words, the results in this paper is also applicable not
only to general relativity but also to any other metric theories
of gravity with general covariance.


Since our motivation of this paper is in higher-order
general-relativistic perturbation theory, readers might think
that the ingredients of this paper is related to the issue
so-called ``linearization
instability''~\cite{Linearization-Instability}. 
However, at least within the level of the ingredients of this
paper, our arguments themselves still have nothing to do with
the issue of linearization instability.
The linearization instability is the issue of the existence of
the solutions to the initial value constraints in the Einstein
equations, while our arguments in this paper is just a
kinematical one and have nothing to do with field equations as
mentioned above. 
Although it will be interesting to reconsider the issue of 
linearization instability from the view point of our
gauge-invariant perturbation theory, this issue is beyond the
current scope of this paper.
To discuss the linearization instability from our view point, it
will be necessary to discuss the above zero-mode problem, at
first.


Although we should take care of the zero-mode problem, we have
almost completed the general framework of the
general-relativistic higher-order gauge-invariant perturbation
theory.
The outline of a proof of Conjecture
\ref{conjecture:decomposition-conjecture} shown in this paper
gives rise to the possibility of the application of our general
framework for the higher-order gauge-invariant perturbation
theory not only to cosmological
perturbations~\cite{kouchan-cosmo-second,kouchan-second-cosmo-matter,kouchan-second-cosmo-consistency}
but also to perturbations of black hole spacetimes or
perturbations of general relativistic stars through an unified
formulation of the general-relativistic perturbation theory. 
Furthermore, as mentioned above, the results in this paper is
also applicable not only to general relativity but also to any
other metric theories of gravity with general covariance.
Therefore, we may say that the wide applications of our
gauge-invariant perturbation theory will be opened due to the
discussions in this paper.
We also leave these development of gauge-invariant perturbation
theories for these background spacetimes as future works.


\section*{Acknowledgments}


The author deeply acknowledged to Professor Robert Manuel Wald
for valuable discussions when the author visited to Chicago
University in 2004.
This work is motivated by the discussions at that time and he
also gave me valuable comments on this manuscript.
The author also thanks to Professor Masa-Katsu Fujimoto in
National Astronomical Observatory of Japan for his various
support and to Professor Misao Sasaki and Professor Takahiro
Tanaka for their hospitality during the workshop on
``Cosmological Perturbation and Cosmic Microwave Background''
(YITP-T-10-05) and their valuable comments.



\appendix
\section{Covariant orthogonal decomposition of symmetric tensors}
\label{sec:S.Deser-1967-J.W.York.Jr-1973-J.W.York.Jr-1974}


Since the each order metric perturbation is regarded as a
symmetric tensor on the background spacetime 
$({\mathcal M}_{0},g_{ab})$ through an appropriate gauge choice, the
covariant decomposition of symmetric tensors is useful and
actually used in the main text. 
Here, we review of the covariant decomposition of symmetric
tensors of the second rank on an curved Riemannian manifold
based on the work by York~\cite{J.W.York-1973-1974}.


On an arbitrary curved Riemannian space $(\Sigma,q_{ab})$
($\dim\Sigma=n$), one can decompose an arbitrary vector or
one-form into its transverse and longitudinal parts as
\begin{eqnarray}
  \label{eq:J.W.York.Jr-1973-1-2-kouchan-1}
  A_{a} = A_{i}(dx^{i})_{a} = \left(D_{i}A_{(L)} + A_{(V)i}\right)(dx^{i})_{a},
  \quad
  D^{i}A_{(V)i} = 0,
\end{eqnarray}
where $D_{i}$ is the covariant derivative associated with the
metric $q_{ab}=q_{ij}(dx^{i})_{a}(dx^{j})_{b}$. 
$A_{(L)}$ is called the longitudinal part or the scalar part and
$A_{(V)i}$ is called the transverse part or vector part of the
vector field $A_{a}$ on $(\Sigma,q_{ab})$, respectively.


Moreover, this decomposition is not only covariant with respect
to arbitrary coordinate transformations, it is also orthogonal
in the natural global scalar product.
To clarify this orthogonality,
York~\cite{J.W.York-1973-1974} introduced the inner
product for the vector fields on $\Sigma$.
This is, for any two vectors $V^{a}$ and $W^{a}$, we have
\begin{eqnarray}
  \label{eq:J.W.York.Jr-1973-1-2}
  \left(V,W\right) := \int_{\Sigma} \epsilon_{q} V^{a}W^{b}q_{ab},
\end{eqnarray}
where $\epsilon_{q}$ denotes the volume element which makes the
integral invariant and the integration extends over the entire
manifold $(\Sigma,q_{ab})$.
In terms of this inner product, the orthogonality of the vector
fields $V^{a}=D^{a}V_{(L)}:=q^{ab}D_{b}V_{(L)}$ and
$W^{a}=q^{ab}V_{(V)b}$ with $D^{a}V_{(V)a}=0$ is given by 
\begin{eqnarray}
  \int_{\Sigma} \epsilon_{q} D_{a}V_{(L)} V_{(V)b} q^{ab}
  = 
  \int_{\partial\Sigma} s_{a} V_{(L)} V_{(V)b} q^{ab}
  -
  \int_{\Sigma} \epsilon_{q} V_{(L)} D_{a}V_{(V)b} q^{ab}
  ,
  \label{eq:J.W.York.Jr-1973-1-2-kouchan-2}
\end{eqnarray}
where $s_{a}$ is the volume element of the $(n-1)$-dimensional
boundary $\partial\Sigma$ of $\Sigma$.
Since the second term of
Eq.~(\ref{eq:J.W.York.Jr-1973-1-2-kouchan-2}) vanishes due to
the condition $D^{a}V_{(V)a}=0$, the inner product $(V,W)$
vanishes if $V_{(L)}$ and $V_{(V)b}$ satisfy some appropriate
boundary conditions at the boundary $\partial\Sigma$ of $\Sigma$
so that the first term of
Eq.~(\ref{eq:J.W.York.Jr-1973-1-2-kouchan-2}) vanishes.
In this sense, the scalar part (the first term in
Eq.~(\ref{eq:J.W.York.Jr-1973-1-2-kouchan-1})) and the vector
part (the second term in
Eq.~(\ref{eq:J.W.York.Jr-1973-1-2-kouchan-1})) orthogonal to
each other.
Geometrically, the decomposition of $1$-forms, and more
generally $p$-forms, leads via de Rham's theorem to a
characterization of topological invariants of $\Sigma$ (i.e.,
Betti Numbers)~\cite{G.DeRham-1960}.


In this appendix, it is assumed that the $n$-dimensional space 
$\Sigma$ is {\it closed} (compact manifolds without boundary)
following York's discussions. 
Although  the decomposition discussed here will also be valid
for the other $n$-dimensional spaces $\Sigma$ with the boundary 
$\partial\Sigma$ with some appropriate boundary conditions at
$\partial\Sigma$, in this Appendix, we choose the closed spaces
as the topology of $\Sigma$ for mathematical convenience.
Through this assumption, in this Appendix, we consider the
TT-decomposition (transverse traceless decomposition) of a
symmetric tensor $\psi^{ab}$ on $\Sigma$, which is defined by 
\begin{eqnarray}
  \label{eq:J.W.York.Jr-1973-2-2}
  \psi^{ab} = \psi_{TT}^{ab} + \psi_{L}^{ab} + \psi_{Tr}^{ab},
\end{eqnarray}
where the longitudinal part is 
\begin{eqnarray}
  \label{eq:J.W.York.Jr-1973-3-2}
  \psi_{L}^{ab} := D^{a}W^{b} + D^{b}W^{a} -
  \frac{2}{n} q^{ab} D_{c}W^{c}
  =: (LW)^{ab}
\end{eqnarray}
and the trace part is
\begin{eqnarray}
  \label{eq:J.W.York.Jr-1973-4-2}
  \psi_{Tr}^{ab} := \frac{1}{n} \psi q^{ab},
  \quad
  \psi := q_{cd} \psi^{cd}.
\end{eqnarray}


Let us suppose that both an arbitrary symmetric tensor field
$\psi^{ab}$ and the metric $q_{ab}$ are $C^{\infty}$ tensor
fields on $\Sigma$.
First, we define $\psi_{TT}^{ab}$ in accordance with
Eq.~(\ref{eq:J.W.York.Jr-1973-2-2}) by
\begin{eqnarray}
  \label{eq:J.W.York.Jr-1973-7-2}
  \psi_{TT}^{ab} := \psi^{ab} - \frac{1}{n} \psi g^{ab} - (LW)^{ab}.
\end{eqnarray}
We note that the tensor $\psi_{TT}^{ab}$ is traceless, i.e., 
\begin{eqnarray}
  \label{eq:J.W.York.Jr-1973-8-2}
  q_{ab}\psi_{TT}^{ab} = 0
\end{eqnarray}
by its construction (\ref{eq:J.W.York.Jr-1973-7-2}).
Further, we require the transversality on the tensor field
$\psi_{TT}^{ab}$, i.e., 
\begin{eqnarray}
  \label{eq:J.W.York.Jr-1973-9-2}
  D_{b}\psi_{TT}^{ab} = 0.
\end{eqnarray}
Equation (\ref{eq:J.W.York.Jr-1973-9-2}) leads to a covariant
equation of the vector field $W^{a}$ in
Eq.~(\ref{eq:J.W.York.Jr-1973-7-2}) as 
\begin{eqnarray}
  \label{eq:J.W.York.Jr-1973-10-2}
  D_{a}(LW)^{ab}
  =
  D_{a}\left(\psi^{ab} - \frac{1}{n} \psi q^{ab}\right).
\end{eqnarray}
The explicit expression of (\ref{eq:J.W.York.Jr-1973-10-2}) is
given by 
\begin{eqnarray}
  \label{eq:J.W.York.Jr-1973-10-2-explicit}
  {\mathcal D}^{bc} W_{c}
  =
  D_{a}\left(\psi^{ab} - \frac{1}{n} \psi q^{ab}\right),
\end{eqnarray}
where the derivative operator ${\mathcal D}^{bc}$ is defined by 
\begin{eqnarray}
  \label{eq:J.W.York.Jr-1973-1-2-kouchan-3}
  {\mathcal D}^{bc}
  :=
  q^{bc}\Delta + \left(1 - \frac{2}{n}\right) D^{b}D^{c} +
  R^{bc}
  , \quad
  \Delta := D^{a}D_{a}
  ,
\end{eqnarray}
where $R^{bc}$ is the Ricci curvature on $(\Sigma,q_{ab})$.


The basic properties of
Eq.~(\ref{eq:J.W.York.Jr-1973-10-2-explicit}) are also discussed
by York~\cite{J.W.York-1973-1974}. 
The operator ${\mathcal D}^{ab}$ defined by
Eq.~(\ref{eq:J.W.York.Jr-1973-1-2-kouchan-3}) is linear and
second order by its definition.
Further, this operator is strongly elliptic, negative-definite,
self-adjoint, and its ``harmonic'' functions are always
orthogonal to the source (right-hand side) in
Eq.~(\ref{eq:J.W.York.Jr-1973-10-2-explicit}). 
Here, ``harmonic'' functions of ${\mathcal D}^{ab}$ means functions
which belong the kernel of the operator ${\mathcal D}^{ab}$.
Moreover, he showed that
Eq.~(\ref{eq:J.W.York.Jr-1973-10-2-explicit}) will always
possess solutions $W^{a}$ which is unique up to conformal Killing
vectors.
Due to this situation, in this paper, we assume that the Green
function $({\mathcal D}^{-1})_{ab}$ defined by 
\begin{eqnarray}
  \label{eq:J.W.York.Jr-1973-kouchan-12}
  ({\mathcal D}^{-1})_{ab}{\mathcal D}^{bc}
  = {\mathcal D}^{bc}({\mathcal D}^{-1})_{ab}
  = \delta_{a}^{\;\;c}
\end{eqnarray}
exists through appropriate boundary conditions at the boundary
$\partial\Sigma$ of $\Sigma$.  
Although York's discussions are for the case of the {\it closed}
space $\Sigma$, we review his discussions here.
In this review, we explicitly write the boundary terms which are
neglected by the closed boundary condition to keep the
extendibility to non-closed $\Sigma$ case of discussions in our
mind.


The ellipticity of an operator depends only upon its 
{\it principal part}, i.e., the highest derivatives acting on
the unknown quantities which it contains.
To see the ellipticity of an operator, we consider the
replacement of the each derivative operator $D_{a}$ occurring in 
its principal part by an arbitrary vector $V_{a}$.
Through this replacement, the principal part of the operator
defines a linear transformation ${\bf \sigma}_{v}$.
The operator is said to be elliptic if ${\bf \sigma}_{v}$ is an
isomorphism~\cite{M.Berger-D.Ebin-1969}.
In the present case, 
\begin{eqnarray}
  \label{eq:J.W.York.Jr-1974-2.4}
  \left[{\bf \sigma}_{v}({\mathcal D})\right]^{ab}
  = V^{b}V^{a} + q^{ab}V_{c}V^{c}.
\end{eqnarray}
Here, ${\bf \sigma}_{v}$ operates on vector $X_{a}$ and defines
a vector-space isomorphism when the determinant of 
${\bf \sigma}_{v}$ is non-vanishing for all non-vanishing
$V^{a}$.
We can verify $\det{\bf \sigma}_{v}\neq 0$, for example, by
choosing $V^{a}=\left(\partial/\partial x^{\mu}\right)^{a}$ in a
local Cartesian frame $\{x^{\mu}\}$.
The operator is said to be {\it strongly elliptic} if all the
eigenvalues of ${\bf \sigma}_{v}$ are nonvanishing and have the 
same sign.
This is easily checked and ${\mathcal D}^{ab}$ is strongly elliptic.


To show that ${\mathcal D}^{ab}$ is negative definite, we consider
the inner product (\ref{eq:J.W.York.Jr-1973-1-2}) of the vector
field ${\mathcal D}W^{a}:={\mathcal D}^{ab}W_{b}$ and $W^{a}$:
\begin{eqnarray}
  \left(W,{\mathcal D}W\right) 
  &=&
  \int_{\Sigma} \epsilon_{q} q_{ab} W^{a}{\mathcal D}^{bc}W_{c}
  \nonumber\\
  &=&
  \int_{\partial\Sigma} s_{c} W_{b} (LW)^{bc}
  - \frac{1}{2}
  \int_{\Sigma} \epsilon_{q}
   (LW)_{bc} (LW)^{bc}
  \label{eq:J.W.York.Jr-1973-13-2}
  ,
\end{eqnarray}
where we use the fact that the tensor $(LW)^{bc}$ is symmetric
and traceless.
Eq.~(\ref{eq:J.W.York.Jr-1973-13-2}) shows that the operator
${\mathcal D}^{ab}$ has the negative eigenvalues in the case where
the first term (boundary term) in
Eq.~(\ref{eq:J.W.York.Jr-1973-13-2}) is neglected, unless
$(LW)^{bc}=0$. 
The self-adjointness of the operator ${\mathcal D}^{ab}$ is follows
from a similar argument in which one integrates by parts twice:
\begin{eqnarray}
  \int_{\Sigma} \epsilon_{q} q_{ab} V^{a}({\mathcal D}W)^{b}
  &=&
  \int_{\partial\Sigma} s_{c} \left[
    V_{b} (LW)^{bc} - \left(LV\right)^{bc}W_{b}
  \right]
  \nonumber\\
  &&
  +
  \int_{\Sigma} \epsilon_{q} W_{b} {\mathcal D}^{bc}V_{c}
  \label{eq:J.W.York.Jr-1973-15-2}
\end{eqnarray}
for any vectors $V$ and $W$, where we use the fact that the
tensor $(LW)^{ab}$ and $(LV)^{ab}$ are symmetric and traceless. 
Eq.~(\ref{eq:J.W.York.Jr-1973-15-2}) shows that the operator
${\mathcal D}^{ab}$ is self-adjoint if the first term (boundary
term) in Eq.~(\ref{eq:J.W.York.Jr-1973-15-2}) is negligible.


When we can neglect the boundary terms in
Eq.~(\ref{eq:J.W.York.Jr-1973-13-2}), the right-hand side of
(\ref{eq:J.W.York.Jr-1973-13-2}) can vanish only if $(LW)^{ab}=0$.
This means either $W^{a}=0$ or $W^{a}$ is a conformal Killing
vector (or Killing vector) of the metric $q_{ab}$.
The condition for a conformal Killing vector is, of course, not
satisfied for an arbitrary metric but this is given by
\begin{eqnarray}
  \label{eq:J.W.York.Jr-1973-16-2}
  {\pounds}_{W}q_{ab} = \lambda q_{ab}
\end{eqnarray}
for some scalar function $\lambda$, where ${\pounds}_{W}$
denotes the Lie derivative with respect to $W^{a}$.
Taking the trace of both sides, we find
\begin{eqnarray}
  \label{eq:J.W.York.Jr-1973-18-2}
  \lambda = \frac{2}{n} D_{c}W^{c}.
\end{eqnarray}
Therefore, $W^{a}$ is a conformal Killing vector if and only if
\begin{eqnarray}
  \label{eq:J.W.York.Jr-1973-19-2}
  D^{a}W^{b} + D^{b}W^{a}
  - \frac{2}{n} q^{ab} D_{c}W^{c}
  \equiv (LW)^{ab} = 0.
\end{eqnarray}
It follows that the only nontrivial solutions of 
${\mathcal D}^{ab}W_{b}=0$ are conformal Killing vectors if they
exist.
Hence the nontrivial ``harmonic'' functions of ${\mathcal D}^{ab}$
are conformal Killing vectors.


We shall now show that even if these ``harmonic'' solutions
exist, they are always orthogonal to the right-hand side of
(\ref{eq:J.W.York.Jr-1973-10-2}) and, hence, can cause no
difficulties in solving equation (\ref{eq:J.W.York.Jr-1973-10-2})
by an eigen function expansion.
Denote the conformal Killing vectors by $W^{a}=C^{a}$, where by
definition $(LC)^{ab}=0$.
Form the scalar product of the right-hand side of
(\ref{eq:J.W.York.Jr-1973-10-2}) with $C$ and integrate by parts
to find
\begin{eqnarray}
  &&
  \int_{\Sigma} \epsilon_{q}
  q_{ac}
  D_{b}\left(\psi^{ab}-\frac{1}{n}q^{ab}\psi\right) C^{c}
  \nonumber\\
  &=&
  \int_{\partial\Sigma} s_{b} \left(\psi^{ab}-\frac{1}{n}q^{ab}\psi\right) C_{a}
  - \frac{1}{2} \int_{\Sigma} \epsilon_{q}
  \left(\psi^{ab}-\frac{1}{n}q^{ab}\psi\right) (LC)_{ab}
  \nonumber\\
  &=&
  0
  ,
  \label{eq:J.W.York.Jr-1973-20-2}
\end{eqnarray}
where we use the fact that $\psi^{ab}-\frac{1}{n}q^{ab}\psi$ is
symmetric and traceless and we also neglect the boundary term.
Hence the source in
Eq.~(\ref{eq:J.W.York.Jr-1973-10-2-explicit}) is in the domain
of $({\mathcal D}^{-1})^{ab}$ and $({\mathcal D}^{-1})^{ab}$ gives the
solution to Eq.~(\ref{eq:J.W.York.Jr-1973-10-2-explicit}) even
in the presence of conformal symmetries.


These results also show that the solution to
Eq.~(\ref{eq:J.W.York.Jr-1973-10-2-explicit}) must be unique up
to conformal Killing vector fields.
Since only $(LW)^{ab}$ enters in the definition
(\ref{eq:J.W.York.Jr-1973-7-2}) of $\psi_{TT}^{ab}$, conformal
Killing vectors cannot affect $\psi_{TT}^{ab}$.


The orthogonality of $\psi_{TT}^{ab}$, $(LW)^{ab}$, and
$\frac{1}{n}\psi q^{ab}$ is easily demonstrated.
We see readily that $\frac{1}{n}\psi q^{ab}$ is pointwise
orthogonal to $(LW)^{ab}$ and to $\psi_{TT}^{ab}$, as
$(LW)^{ab}$ and $\psi^{ab}_{TT}$ are both trace-free.
To show that $\psi_{TT}^{ab}$ and $(LV)^{ab}$ are orthogonal for
any vector $V^{a}$ and any TT tensor, we only show that 
\begin{eqnarray}
  \int_{\Sigma} \epsilon_{q} q_{ac}q_{bd} (LW)^{ab}\psi_{TT}^{cd}
  &=&
  \int_{\partial\Sigma} s_{a}\left(2 W_{b}\psi_{TT}^{ab}\right)
  - \int_{\Sigma} \epsilon_{q} \left(
    2 W_{b} D_{a}\psi_{TT}^{ab}
  \right)
  \nonumber\\
  &=&
  0
  ,
  \label{eq:J.W.York.Jr-1973-20.5}
\end{eqnarray}
where we use the fact that the tensor $\psi_{TT}^{ab}$ is
symmetric, traceless, and transverse
(\ref{eq:J.W.York.Jr-1973-9-2}).
We also neglect the boundary term in
Eq.~(\ref{eq:J.W.York.Jr-1973-20.5}).
Thus, we conclude that the decomposition defined by
(\ref{eq:J.W.York.Jr-1973-7-2}) exists, is unique, and is
orthogonal.


One can further decompose the vector $W^{a}$ uniquely into 
its transverse and longitudinal parts with respect to the metric
$q_{ab}$. 
This splitting is orthogonal, as in
Eq.~(\ref{eq:J.W.York.Jr-1973-1-2-kouchan-1}).


Since the above discussions are for the case of a closed spaces
$\Sigma$, careful discussions on the boundary terms, which are
neglected in the case of a closed $\Sigma$, is necessary if we
extend the above arguments to the case of a non-closed $\Sigma$.
However, we do not go into these detailed issues.
Instead, in the main text, we assume that the existence of the
Green function of the derivative operator ${\mathcal D}^{ab}$ and
use the transverse-traceless decomposition for an arbitrary
symmetric tensor on $\Sigma$ discussed here.



\begin{thebibliography}{9}
\bibitem{H.Stephani-D.Kramer-M.A.N.MacCallum-C.Hoenselaers-E.Herlt-2003}
  H.~Stephani, D.~Kramer, M.~A.~N.~MacCallum, C.~Hoenselaers,
  E.~Herlt, {\it Exact solutions of Einstein's Field Equations
    Second Edition}, Cambridge Monographs on Mathematical
  Physics (Cambridge University Press, 2003).
\bibitem{Bardeen-1980}
  J.~M.~Bardeen, Phys.~Rev.\ D\ {\bf 22} (1980), 1882;
  H.~Kodama and M.~Sasaki, Prog.~Theor.~Phys.~Suppl.\ No.~78 (1984), 1;
  V.~F.~Mukhanov, H.~A.~Feldman and R.~H.~Brandenberger,
  Phys.~Rep.\ {\bf 215} (1992), 203.
\bibitem{Gerlach_Sengupta-1979}
  U.H.~Gerlach and U.K.~Sengupta, Phys.~Rev.\ D\ {\bf 19} (1979), 2268;
                                  Phys.~Rev.\ D\ {\bf 20} (1979), 3009;
                                  Phys.~Rev.\ D\ {\bf 22} (1980), 1300;
                                  J. Math. Phys.\ {\bf 20} (1979), 2540;
  C.~Gundlach and J.M.~Mart\'in-Garc\'ia, Phys.~Rev.\ D{\bf 61}
  (2000), 084024;
  J.M.~Mart\'in-Garc\'ia and C.~Gundlach, Phys.~Rev.\ D{\bf 64}
  (2001), 024012;
  S.~Chandrasekhar, {\it The mathematical theory of black holes} (Oxford:
  Clarendon Press, 1983).
\bibitem{Kodama-Ishibashi-2004}
  H.~Kodama, A.~Ishibashi and O.~Seto, Phys. Rev. D {\bf 62} (2000), 064022;
  S.~Mukohyama, Phys. Rev. D {\bf 62} (2000), 084015;
  H.~Kodama and A.~Ishibashi, Prog. Theor. Phys. {\bf 110} (2004), 701;
                              Prog. Theor. Phys. {\bf 111} (2004), 29. 
\bibitem{Tomita-1967-Non-Gaussianity}
  K.~Tomita, Prog. Theor. Phys. {\bf 37} (1967), 831;
             Prog. Theor. Phys. {\bf 45} (1971), 1747;
             Prog. Theor. Phys. {\bf 47} (1972), 416;
  V.~Acquaviva, N.~Bartolo, S.~Matarrese, and A.~Riotto,
  Nucl. Phys. B {\bf 667} (2003), 119;
  J.~Maldacena, J. High Energy Phys.\ {\bf 05} (2003), 013;
  K.~A.~Malik and D.~Wands, Class. Quantum Grav. {\bf 21} (2004), L65;
  N.~Bartolo, S.~Matarrese and A.~Riotto, Phys. Rev. D {\bf 69}
  (2004), 043503;
  J. High Energy Phys.\ {\bf 04} (2004), 006;
  D.~H.~Lyth and Y.~Rodr\'iguez, Phys. Rev. D {\bf 71} (2005), 123508;
  F.~Vernizzi, Phys. Rev. D {\bf 71} (2005), 061301(R);
  S.~Matarrese, S.~Mollerach and M.~Bruni,
  Phys. Rev. D {\bf 58} (1998), 043504;
  N.~Bartolo, S.~Matarrese and A.~Riotto,
  J. Cosmol. Astropart. Phys.\ {\bf 01} (2004), 003;
  Phys. Rev. Lett. {\bf 93} (2004), 231301;
  N.~Bartolo, E.~Komatsu, S.~Matarrese and A.~Riotto,
  Phys. Rep. {\bf 402} (2004), 103;
  N.~Bartolo, S.~Matarrese and A.~Riotto, JCAP {\bf 0605} (2006), 010;
  K.~Tomita, Phys. Rev. D {\bf 71} (2005), 083504;
             Phys. Rev. D {\bf 72} (2005), 103506;
             Phys. Rev. D {\bf 72} (2005), 043526;
  K.~A.~Malik and D.~Wands,
  Phys.\ Rept.\  {\bf 475}, 1 (2009);
  C.~Pitrou, Class. Quantum Grav. {\bf 24} (2007), 6127;
  {\it ibid}. {\bf 26} (2009), 065006.
\bibitem{M.Bruni-S.Soonego-CQG1997}
  M.~Bruni, S.~Matarrese, S.~Mollerach and S.~Sonego,
  Class. Quantum Grav.\ {\bf 14} (1997), 2585;
  M.~Bruni and S.~Sonego, Class.~Quantum~Grav.\ {\bf 16} (1999), L29.
\bibitem{S.Sonego-M.Bruni-CMP1998}
  S.~Sonego and M.~Bruni, Commun.~Math.~Phys.\ {\bf 193} (1998), 209.
\bibitem{Nakamura:2010yg}
  K.~Nakamura, Advances in Astronomy, {\bf 2010} (2010), 576273.
\bibitem{kouchan-cosmo-second}
  K.~Nakamura, Phys. Rev. D {\bf 74} (2006), 101301(R);
  K.~Nakamura, Prog. Theor. Phys. {\bf 117} (2007), 17.
\bibitem{Non-Gaussianity-observation}
  E.~Komatsu et al., Astrophys. J. Suppl. Ser. {\bf 180} (2009), 330;
                    arXiv:1001.4538[astro-ph.CO].
\bibitem{Gleiser-Nicasio-2000}
  M.~Campanelli, C.~O.~Lousto, Phys. Rev. D {\bf 59} (1999),
  124022; 
  R.~J.~Gleiser, C.~O.~Nicasio, R.~H.~Price and J.~Pullin,
  Phys.~Rep.\ {\bf 325} (2000), 41, and references therein.
\bibitem{Kojima-1997}
  Y.~Kojima, Prog.~Theor.~Phys.~Suppl.\ No.128 (1997), 251;
  A.~Passamonti, M.~Bruni, L.~Gualtieri and C.F.~Sopuerta,
  Phys. Rev. D {\bf 71} (2005), 024022.
\bibitem{kouchan-gauge-inv}
  K.~Nakamura, Prog.~Theor.~Phys. {\bf 110}, (2003), 723.
\bibitem{kouchan-second} 
  K.~Nakamura, Prog. Theor. Phys. {\bf 113} (2005), 481.
\bibitem{kouchan-papers}
  K.~Nakamura, A.~Ishibashi and H.~Ishihara, Phys.~Rev.\ D{\bf 62} (2000),
  101502(R);
  K.~Nakamura and H.~Ishihara, Phys.~Rev.\ D\ {\bf 63} (2001),
  127501;
  K.~Nakamura, Class.~Quantum~Grav.\ {\bf 19} (2002), 783;
               Phys.~Rev.\ D\ {\bf 66} (2002), 084005;
               Prog.~Theor.~Phys. {\bf 110}, (2003), 201.
\bibitem{kouchan-second-cosmo-matter}
  K.~Nakamura, Phys. Rev. D {\bf 80} (2009), 124021.
\bibitem{kouchan-second-cosmo-consistency}
  K.~Nakamura, Prog. Theor. Phys. {\bf 121} (2009), 1321.
\bibitem{kouchan-decomp-letter-version}
  K.~Nakamura, Class.~Quantum~Grav.\ {\bf 28} (2011), 122001.
\bibitem{Wald-book}
  R.~M.~Wald, {\it General Relativity} (Chicago, IL, University of
  Chicago Press, 1984).
\bibitem{R.K.Sachs-1964}
  R.~K.~Sachs, ``Gravitational Radiation'', in {\it Relativity,
    Groups and Topology} ed. C.~DeWitt and B.~DeWitt, (New York: 
  Gordon and Breach, 1964). 
\bibitem{J.M.Stewart-M.Walker11974}
  J.~M.~Stewart and M.~Walker, Proc.~R.~Soc.~London\ A {\bf 341}
  (1974), 49;
  J.~M.~Stewart, Class.~Quantum~Grav.\ {\bf 7} (1990), 1169;
                 {\it Advanced General Relativity} (Cambridge
                 University Press, Cambridge, 1991). 
\bibitem{Pereira:2007yy}
  T.~S.~Pereira, C.~Pitrou and J.~P.~Uzan,
  JCAP {\bf 0709}, 006 (2007);
  C.~Pitrou, T.~S.~Pereira and J.~P.~Uzan,
  JCAP {\bf 0804}, 004 (2008).
\bibitem{S.Deser-1967}
  S.~Deser, Ann. Inst. H. Poincar\'e {\bf 7} (1967), 149.
\bibitem{Christopherson-Malik-Matravers-Nakamura-comparison}
  A.~J.~Christopherson, K.~A.~Malik, D.~R.~-Matravers,
  K.~Nakamura, Class.~Quantum~Grav.\ {\bf 28} (2011), 225024.
\bibitem{C.Uggla-J.Wainwright-2011}
  C.~Uggla and J.~Wainwright, Class.~Quantum~Grav.\ {\bf 28}
  (2011), 175017. 
\bibitem{J.W.York-1973-1974}
  J.~W.~York, Jr. J.~Math.~Phys. {\bf 14} (1973), 456;
                  Ann. Inst. H. Poincar\'e {\bf 21} (1974), 319.
\bibitem{Linearization-Instability}
  A.~E.~Fisher and J.~E.~Marsden, Bull. Am. Soc. {\bf 79} (1973),
  997; 
  D.~Brill and S.~Deser, Commun. Math. Phys. {\bf 32} (1973),
  291; 
  V.~Moncrief, J. Math. Phys. {\bf 16} (1975), 493;
  V.~Moncrief, J. Math. Phys. {\bf 17} (1976), 1893;
  V.~Moncrief, Phys. Rev. D {\bf 18} (1978), 983;
  A.~E.~Fischer, J.~E.~Marsden, and V.~Moncrief,
  Ann. Inst. Henri Poincar\'e, {\bf 33} (1980), 147; 
  J.~M.~Arms, J.~E.~Marsden, and V.~Moncrief,
  Commun. Math. Phys. {\bf 78} (1981), 455;
  J.~M.~Arms, J.~E.~Marsden, and V.~Moncrief,
  Annals of Physics {\bf 144} (1982), 81;
  A.~Higuchi, Class.~Quantum~Grav.\ {\bf 8} (1991), 1961; 
  B.~Losic and W.~G.~Unruh, Phys. Rev. D {\bf 74} (2006),
  023511; 
  B.~Losic and W.~G.~Unruh, Phys. Rev. Lett. {\bf 101} (2008),
  111101.
\bibitem{G.DeRham-1960}
  G.~De~Rham, ``vari\'et\'es Diff\'erentiables,'' Hermann,
  Paris, 1960.
\bibitem{M.Berger-D.Ebin-1969}
  M.~Berger and D.~Ebin, J. Diff. Geom. {\bf 3} (1969), 379.
\bibitem{K.Nakamura-IJMPD-2012}
  K.~Nakamura, Int. J. Mod. Phys. D {\bf 21} (2012), 124004.
\end{thebibliography}
\end{document}